\documentclass{aastex63}
\usepackage{bm}
\usepackage{ulem}
\usepackage{amsmath}
\usepackage{verbatim} 

\usepackage[utf8]{inputenc}
\usepackage[T1]{fontenc}

\usepackage[encapsulated]{CJK}

\usepackage{rotating} 
\usepackage{lineno} 

\usepackage{xcolor} 
\usepackage{ulem} 

\received{XXX}
\revised{YYY}
\accepted{ZZZ}

\submitjournal{ApJ}

\shorttitle{ZTF Spider Pulsar Survey}
\shortauthors{Lu et al.}
\graphicspath{{./}{figures/}}

\begin{document}
\begin{CJK*}{UTF8}{gbsn}


\title{A systematic search for redback and black widow candidates based on the 4FGL-DR3 unassociated sources and the Zwicky Transient Facility data}

\correspondingauthor{P. H. T. Tam, Liangliang Ren}
\email{tanbxuan@mail.sysu.edu.cn, rll@ahstu.edu.cn}

\author[0009-0006-4992-1881]{Chunyan Lu (卢春燕)}
\affiliation{School of Physics and Astronomy, Sun Yat-sen University, Zhuhai 519082, China}
\affiliation{CSST Science Center for the Guangdong-Hongkong-Macau Greater Bay Area, Sun Yat-sen University, Zhuhai 519082, China}

\author[0000-0002-1428-4003]{Liangliang Ren (任亮亮)}
\affiliation{School of Electrical and Electronic Engineering, Anhui Science and Technology University, Bengbu, Anhui 233030, China}

\author[0009-0008-9942-620X]{Jiamao Lin (林佳茂)}
\affiliation{School of Physics and Astronomy, Sun Yat-sen University, Zhuhai 519082, China}
\affiliation{CSST Science Center for the Guangdong-Hongkong-Macau Greater Bay Area, Sun Yat-sen University, Zhuhai 519082, China}

\author[0000-0001-5546-8549]{Wenjun Huang (黄文俊)}
\affiliation{School of Physics and Astronomy, Sun Yat-sen University, Zhuhai 519082, China}
\affiliation{CSST Science Center for the Guangdong-Hongkong-Macau Greater Bay Area, Sun Yat-sen University, Zhuhai 519082, China}

\author{Hewen Yang (杨何文)}
\affiliation{School of Physics and Astronomy, Sun Yat-sen University, Zhuhai 519082, China}
\affiliation{CSST Science Center for the Guangdong-Hongkong-Macau Greater Bay Area, Sun Yat-sen University, Zhuhai 519082, China}

\author[0000-0002-1262-7375]{P. H. Thomas Tam}
\affiliation{School of Physics and Astronomy, Sun Yat-sen University, Zhuhai 519082, China}
\affiliation{CSST Science Center for the Guangdong-Hongkong-Macau Greater Bay Area, Sun Yat-sen University, Zhuhai 519082, China}



\begin{abstract}

Spider pulsars constitute a distinct subset within the domain of radio millisecond pulsars, divided further into the categories of black widows and redbacks. Evident across multiple wavelengths, these pulsars manifest periodic variations and reside within binary systems. Investigating and discovering additional spider-type pulsars carries significant implications for comprehending the evolution of high-mass stars. Particularly crucial is the validation of the "Recycling" theory of millisecond pulsar genesis. In this investigation, we systematically explore spider pulsar binary systems utilizing time-domain variability data from the \emph{Zwicky Transient Facility}, in conjunction with Fermi unassociated gamma-ray sources sourced from the 4FGL-DR3 catalog. We have implemented a time-domain data processing pipeline utilizing the Lomb-Scargle Periodogram algorithm, integrated with the wget data crawling technology. This approach has led to the identification of 194 ellipsoidal variables and irradiation-type binary stars. Subsequent refinement through the Gaia Hertzsprung-Russell diagram has culled a selection of 24 spider pulsar gold sample candidates. By incorporating the 4FGL 95\% confidence error ellipse, the pool was narrowed down to 19 gold sample candidates. Utilizing the Gaia color-reduced proper motion diagram further refined the selection to 9 gold sample candidates. These newly identified spider pulsar candidates will inform subsequent observational campaigns across radio, X-ray, and optical spectroscopy, thereby facilitating a deeper validation of their physical characteristics.

\end{abstract}



\section{Introduction} \label{sec:intro}

Millisecond pulsars (MSPs) are a special population of radio pulsars characterized by their rapid rotation periods (typically spin periods $\rm P_{s} \lesssim $ 30 ms), spin period derivatives ranging from $10^{-21}$ to $10^{-18}$ $\rm s~s^{-1}$, magnetic field strengths B $\sim 10^{7} - 10^{9}$ G, and characteristic ages spanning from $10^{7}$ to $10^{11}$ years \citep{Camilo1994,Lorimer2008}. 
According to the standard ``recycling" scenario, MSPs are spun up to rapid spin periods through the accretion of matter from a companion star 
\citep{Alpar1982,Radhakrishnan1982,Jiang2015,Baglio2016,Chugunov2017}. This means they need to form in binary systems, as otherwise, an isolated pulsar would enter the so-called ``graveyard" zone in the period-period derivative ($\rm P-\dot P$) diagram \citep{Liu2022}.

Spider pulsars are a specialized sub-group within the population of MSPs, and they can be further divided into two sub-classes \citep{Hui2019,Polzin2020}: black widows (BWs) and redbacks (RBs). BWs are composed of a pulsar and a low-mass degenerate companion star with masses $\rm M \lesssim~0.1$ $\rm M_{\odot}$. RBs are closely related to BWs but differ in that they consist of a pulsar and a late-type non-degenerate star, with companion masses ranging from 0.2 to 1.0 $\rm M_{\odot}$.

The search and confirmation of new spider MSPs play a key role in a range of crucial astrophysical studies, including understanding the physical mechanisms behind the formation of MSPs through the ``recycling" scenario \citep{Baglio2016,Chugunov2017}, investigating mass transfer in ultra-compact binary systems \citep{Chen2013}, exploring the acceleration of high-energy particles in MSP winds \citep{Takata2014,Campana2016,Hui2018,Hui2019,Veledina2019}, and accurately measuring the pulse arrival times of MSPs in the Pulsar Timing Array (PTA) to detect the stochastic gravitational wave background \citep{Verbiest2009,Arzoumanian2018,Arzoumanian2020,Alam2021a,Alam2021b}.

Due to the efficient searches conducted by radio telescopes such as the Australia Telescope National Facility (ATNF) \citep{Manchester2005} and Five-hundred-meter Aperture Spherical radio Telescope (FAST) \citep{Nan2011}, the ATNF catalog contains 3,389 radio pulsars\footnote{\url{https://www.atnf.csiro.au/research/pulsar/psrcat/expert.html}}，and the FAST Galactic Plane Pulsar Snapshot (GPPS) survey has discovered over 566 pulsars\citep{Han2021,Zhou2023,Su2023}. In total, over 512 radio pulsars have been identified as belonging to the MSP population \citep{Bhattacharyya2022}.
The Fermi Large Area Telescope (Fermi-LAT) has detected 278 gamma-ray pulsars \citep{Abdo2010,Abdo2013,Abdollahi2022}, with over half (128/278) being MSPs (118 radio selected and 10 gamma-ray selected).
The known spider MSPs include 44 BWs and 26 RBs, which have been confirmed in the galactic field or globular clusters\citep{Hui2019,Lee2023}.

The deep radio search for Fermi gamma-ray sources is expected to hunt for more MSPs, especially for high energetic spider pulsars \citep{2021SCPMA..6429562W,2023MNRAS.519.5590C,2024arXiv240209366B}. Generally, the discovery of a radio MSP is the ultimate way to confirm a new spider system. Once a MSP is discovered, its radio pulse period can accurately aid in folding the gamma-ray pulses. Meanwhile, subsequent radio timing can further constrain the Keplerian parameters of the binary system. In comparison to the gamma-ray luminosity  which is typically proportional to spin-down luminosity ($\dot{E}$), the radio luminosity of spider exhibits only a very weak coupling with $\dot{E}$, and the faction of radio luminosity to $\dot{E}$ is also lower \citep{2024arXiv240209366B}. However, not all spider pulsars are easily detectable in radio pulses. Besides the possibility of being radio faint, various factors may affect their detectability: spider may be eclipsed by the material stripped from the companion, leading to radio pulsations undetectable at low frequency for a significant fraction ($\lesssim$ 50\%) of the orbital period, due to the rapid rise of dispersion measure \citep{2024arXiv240209366B}; shorter spin period demands higher temporal resolution, and shorter orbital period requires larger acceleration for acceleration searches; for redback systems, a low accretion state may happen resulting in non-detection of radio pulses \citep{2014ApJ...790...39S}.
 
Among the known spider MSPs, periodic variations in the optical observations have been discovered \citep{Swihart2021,Swihart2022,Halpern2022a,Halpern2022b,Au2023}, and there are two main causes for these optical periodicity: pulsar irradiation and ellipsoidal modulation. The former effect is due to the fact that the side of the companion facing the pulsar is hotter and thus brighter, while the latter is attributed to tidal distortion of the companion.
Based on follow-up observations combining the Fermi LAT catalog with optical and X-ray periodicity variations, more than a dozen spider pulsars with orbital periods less than 1 day have been confirmed \citep{Braglia2020}. 
In particular, a search utilizing archived data from the Transiting Exoplanet Survey Satellite (TESS) for known pulsars or candidates has successfully identified five objects exhibiting orbital periodic variations, thereby validating the feasibility of detecting pulsar binaries through time-domain photometric data \citep{Pal2020}.

The 4FGL-DR3 catalog is based on more than 13 years of the full-sky gamma-ray data collected by the Fermi-LAT satellite \citep{Abdollahi2022}. Among a total of 6658 gamma-ray sources, it includes approximately 2157 unidentified or unassociated sources, indicating the presence of unknown counterparts in other electromagnetic wavelengths. The Zwicky Transient Facility (ZTF), as one of the highly efficient facilities for sky surveys in the northern hemisphere, provides time-domain photometric data for billions of objects. It has achieved significant success in searching for short period black widow binary \citep{Burdge2022}, LISA gravitational wave sources \citep{Burdge2019,Burdge2020}, and close white dwarf binaries \citep{Ren2023}.

In this work, we conducted a systematic search for optical counterparts of spider binary candidates by utilizing the archival time-domain data from the Zwicky Transient Facility (ZTF) in conjunction with the sample of unidentified or unassociated gamma-ray sources from the 4FGL-DR3 catalog.
The structure of the paper is outlined as follows: In Section \ref{sec:methods}, we introduce the sample selection of spider pulsars and the methods for processing their light curves. In Section \ref{sec:analaysisresults}, we describe the results of data processing and the screening scheme for the gold sample of spider pulsar candidates. In Section \ref{sec:discussion}, we present the screening results of the gold sample in other databases and discuss the period-optical luminosity relationship for spider pulsar candidates, and Section \ref{sec:conclusion} is the final conclusion.

\begin{figure*}[htpb]
\begin{center}
\includegraphics[width=0.9\textwidth]{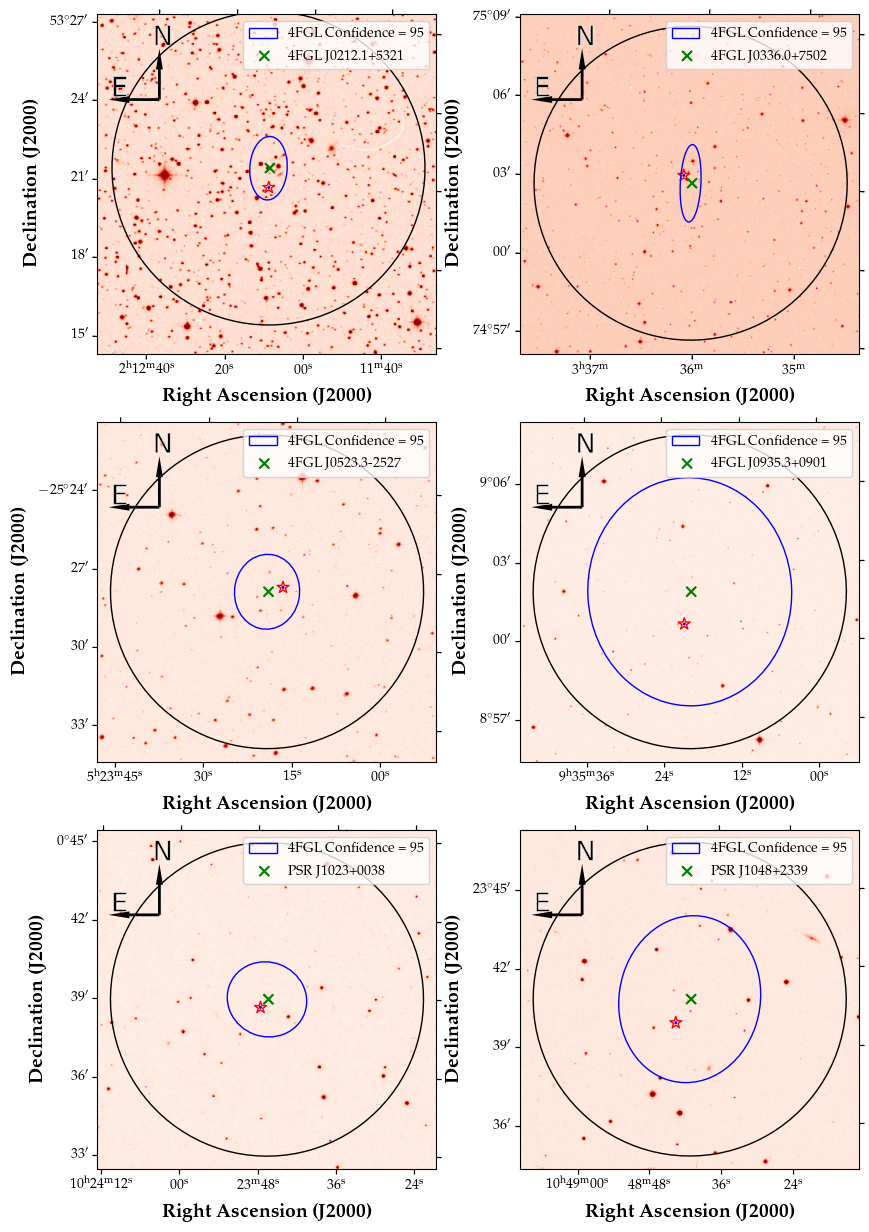}
\caption{The sky maps of the 12 known millisecond pulsars in the test sample. The blue ellipses represent the 4FGL 95\% confidence error ellipses, the black circles indicate the search area within 6 arcminutes in the ZTF variability data. The green crosses mark the central coordinates of the Fermi gamma-ray sources, and the red pentagons denote the optical counterparts in the ZTF. The background data for the sky maps are sourced from the POSS-1 red band image, accessed through the Virtual Observatory archives.}
\label{fig:possimage1}
\end{center}
\end{figure*}

\begin{figure*}[htpb]
\begin{center}
\includegraphics[width=0.9\textwidth]{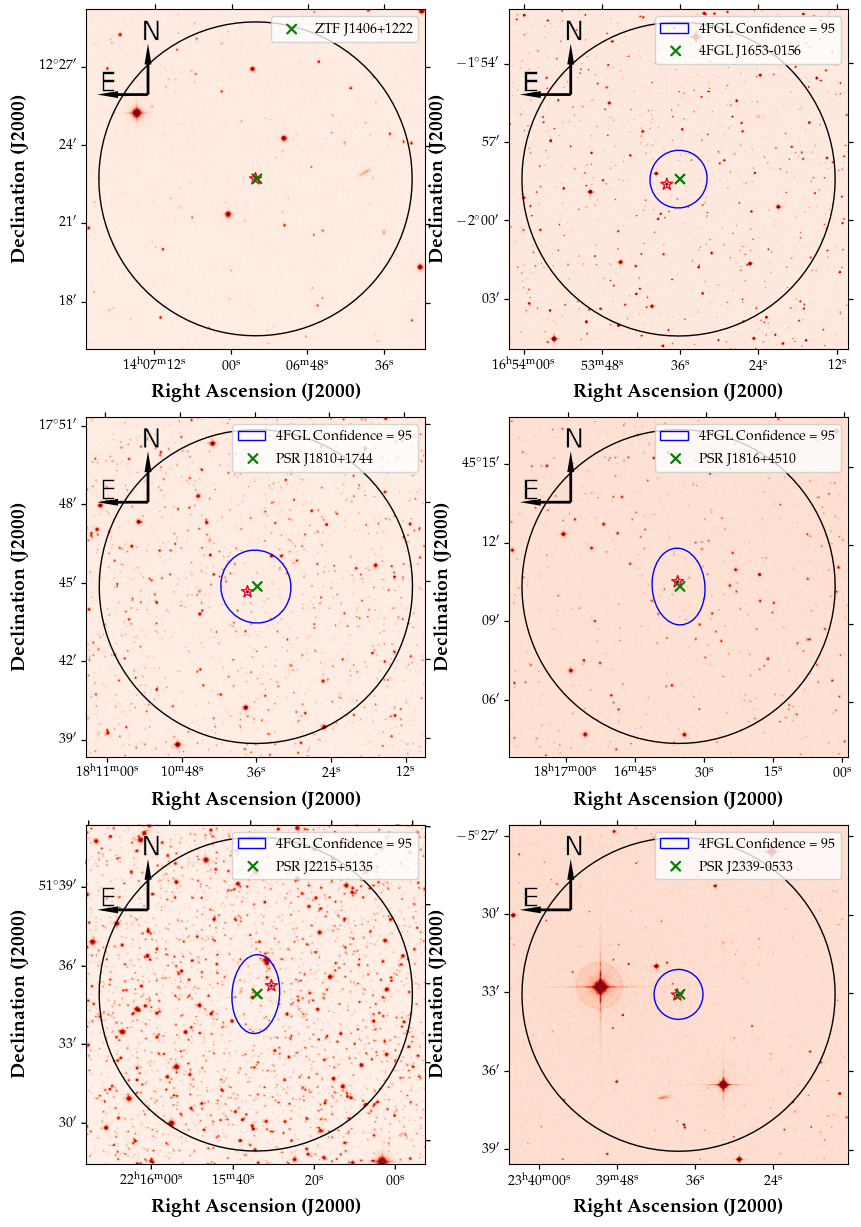}
\caption{POSS-1(red) Image of the spider test sample. Same as in Figure \ref{fig:possimage1} (continued).}
\label{fig:possimage2}
\end{center}
\end{figure*}

\begin{deluxetable*}{ccccccc}
\tablenum{1}
\tablecaption{Spider pulsar binaries with confirmed periodic variations in the ZTF data were identified from the test sample.  \label{tab:knowspiders}}
\tablewidth{0pt}
\tablehead{
\colhead{Name} & \colhead{Right Ascension}  & \colhead{Declination} &
\colhead{Orbital period} & \colhead{ Dominant types of } & \colhead{Type} & \colhead{Ref.}\\
\colhead{} & \colhead{(deg.)} & \colhead{(deg.)}  & \colhead{(min)} &
\colhead{photometric variability } & \colhead{} & \colhead{}
}
\startdata
PSR J0212+5321 & 33.043633   & +53.360822 & 1252.152 & Ellipsoidal & Redback  
 & 1,2,3,4 \\
4FGL J0336.0+7502 & 54.042790   & +75.054776 & 223.09068 & Irradiation &  Black Widow & 5 \\
3FGL J0523.5-2529 &  80.820566  & -25.460282 & 990.917 & Ellipsoidal & Redback & 4,6,7  \\
4FGL J0935.3+0901 & 143.836305 & 9.009949 &  146. 207 & Irradiation & Black Widow candidate & 8,9 \\
PSR J1023+0038 &  155.948717  & +0.644668 & 570.500 & Ellipsoidal & Redback & 10,11,12 \\
PSR J1048+2339 &  162.180978  & +23.664910 & 360.747 & Irradiation & Redback & 13  \\
ZTF J1406+1222 &  211.734045  & +12.378832 & 62.002 & Irradiation & Black Widow  & 14 \\
PSR J1653-0158 &  253.408588  & -1.976984 & 74.8005 & Irradiation & Black Widow & 15,16 \\
PSR J1810+1744 &  272.655333  & +17.743717 & 213.365 & Irradiation & Black Widow  & 17,18 \\
PSR J1816+4510 &  274.149622  & +45.176085 & 519.687 & Ellipsoidal & Redback/Black Widow & 19,20 \\
PSR J2215+5135 &  333.886238  & +51.593481 & 248.4045 & Irradiation & Black Widow & 21,22 \\
PSR J2339-0533 &  354.911313  & -5.551317 & 278.062 & Irradiation & Black Widow & 23,24,25 \\
\enddata
\tablecomments{The coordinates (J2000.0) and basic photometric characteristics of the known spider Pulsar binaries in twelve test samples were confirmed using ZTF Data validation.
\tablerefs{(1) \citet{Li2016}; (2) \citet{Linares2017}; (3) \citet{Shahbaz2017}; (4) \citet{Pal2020}; (5) \citet{Li2021}; (6) \citet{Xing2014}; (7) \citet{Halpern2022b}; (8) \citet{Wang2020}; (9) \citet{Halpern2022a}; (10) \citet{Illiano2023}; (11) \citet{Shahbaz2019}; (12) \citet{Miraval2022}; (13) \citet{Deneva2016}; (14) \citet{Burdge2022};  (15) \citet{Nieder2020}; (16) \citet{Long2022}; (17) \citet{Gentile2014}; (18) \citet{Polzin2018}; (19) \citet{Kaplan2012}; (20) \citet{Polzin2020}; (21) \citet{Romani2015}; (22) \citet{Voisin2020}; (23) \citet{Romani2011}; (24) \citet{An2020}; (25) \citet{Kandel2020}}}
\end{deluxetable*}

\section{Methods} \label{sec:methods}
\subsection{Sample Selection}
\subsubsection{Test Sample}

As a pre-search validation, we initially selected a set of known spider MSPs from recent studies as test samples. 
The purpose of these test samples was to verify the reliability of our light curve processing pipeline code. The reliability of using ZTF (or PTF) and TESS data  to search for periodic variations in these test sources and other Fermi Redbacks has been thoroughly demonstrated in previous studies: \citep{Bellm2016} (PSR J2129-0429); \citep{Deneva2016} (PSR J1048+2339); \citep{Halpern2022a} (4FGL J0935.3+0901); \citep{Halpern2022b} (3FGL J0523.5-2529); \citep{Karpova2023} (4FGL J2054.2+6904); \citep{Perez2023} (PSR J0212+5320) and \citep{Pal2020}.

The test samples were derived from various sources: 43 black widows (BW-Field) and redbacks (RB-Field) in the Galactic field, and 12 spider pulsar candidates from Table 3 curated by \citet{Hui2019}, 5 Redback Systems (Candidates) compiled by \citet{Pal2020}, 48 MSP candidates collected by \citet{Braglia2020}, 278 sources from the Fermi-LAT Detected Gamma-Ray Pulsars list \footnote{\url{https://confluence.slac.stanford.edu/display/GLAMCOG/Public+List+of+LAT-Detected+Gamma-Ray+Pulsars}}, and recently optically identified Fermi-LAT unassociated sources \citep{Halpern2022b,Swihart2022,Braglia2020,VanStaden2022,Swihart2022a}. We conducted a reorganization and classification of these samples, taking into account the observation coverage of ZTF in regions with a $\rm decl. > -28^{\circ}$. The final test sample comprised a total of 99 sources, including 42 RB and BW systems, as well as 57 MSPs.

\begin{figure*}[htpb]
\begin{center}
\includegraphics[width=1.0\textwidth]{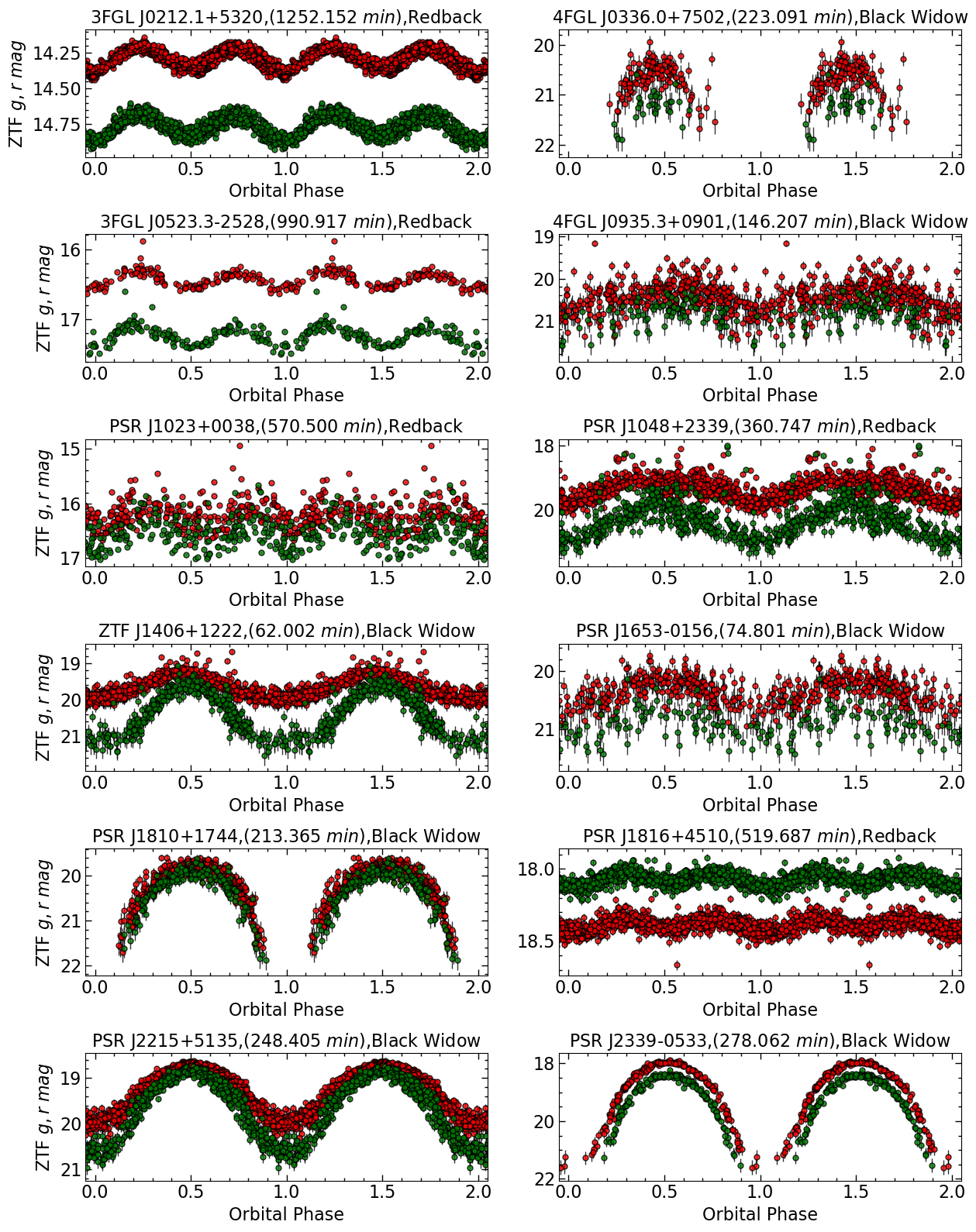}
\caption{Phase-folded ZTF light curves for the 12 known millisecond pulsars within the test sample. The red data points correspond to the ZTF-r band, while the green data points correspond to the ZTF-g band.}
\label{fig:testsample}
\end{center}
\end{figure*}

\begin{figure*}[htpb]
\begin{center}
\includegraphics[width=1.0\textwidth]{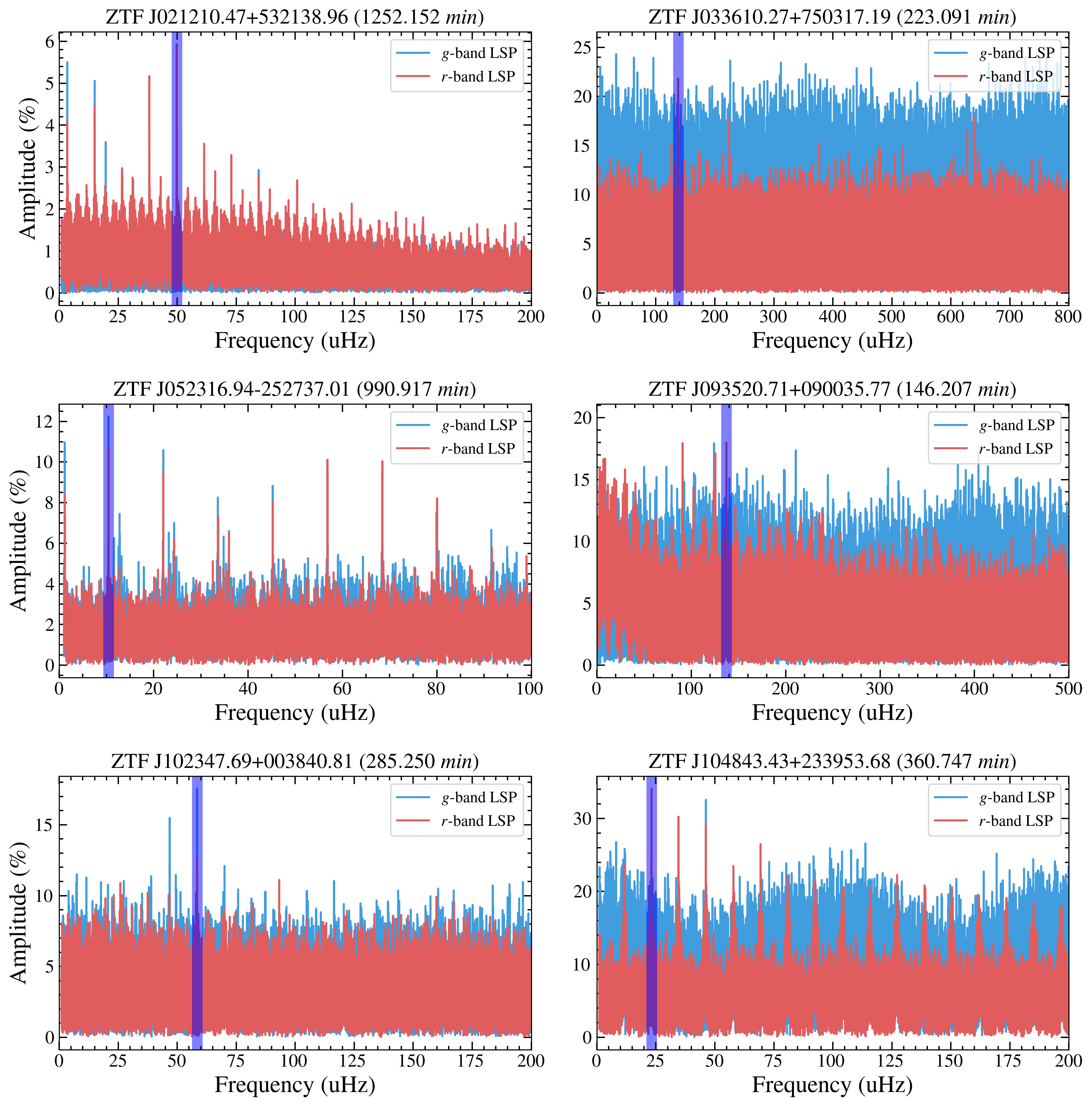}
\caption{The Lomb-Scargle (LS) periodograms of the frequency spectra for the 12 known millisecond pulsars in the test sample. The red line represents the frequency spectrum for the ZTF-r band, while the blue line represents the frequency spectrum for the ZTF-g band. The blue rectangle indicates the frequency corresponding to the best period of variability for the pulsar binary system.}
\label{fig:LSPSD1}
\end{center}
\end{figure*}

\begin{figure*}[htpb]
\begin{center}
\includegraphics[width=1.0\textwidth]{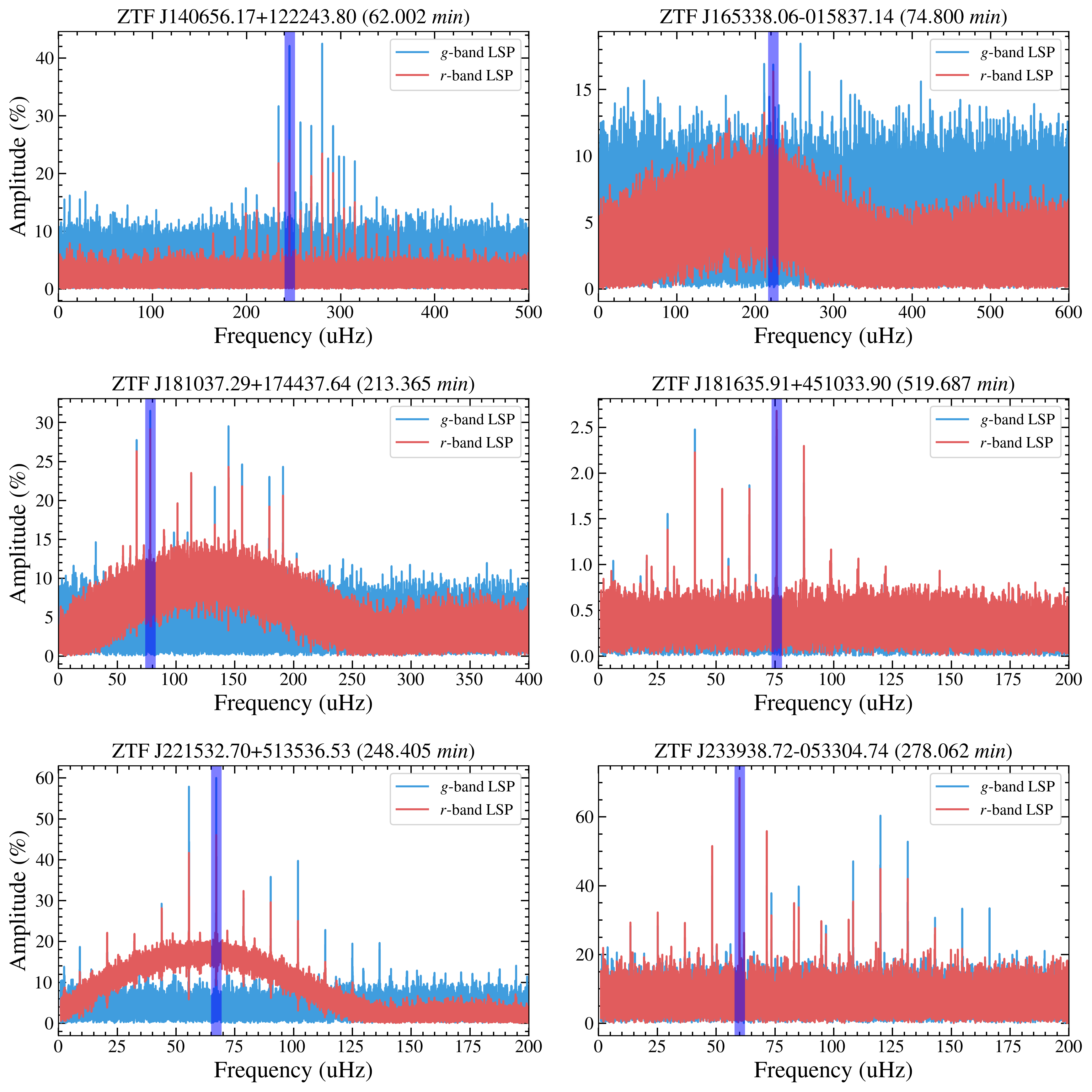}
\caption{The Lomb-Scargle (LS) periodograms of the frequency spectra for the 12 known millisecond pulsars in the test sample. Same as in Figure \ref{fig:LSPSD1} (continued).}
\label{fig:LSPSD2}
\end{center}
\end{figure*}

\subsubsection{Search Sample}

The 4FGL-DR3 catalog \footnote{\url{https://fermi.gsfc.nasa.gov/ssc/data/access/lat/12yr_catalog/}} is a compilation of scientific data accumulated by the Fermi satellite over a period of 12 years, covering the energy range from 50 MeV to 1 TeV \citep{Abdollahi2022}. It includes 6658 sources, with an addition of 1607 sources compared to DR1 and 900 sources compared to DR2 \citep{Abdollahi2020}. In the DR3 catalog, 120 millisecond pulsars have been confirmed through pulsations, and 35 sources (without pulsations) have been associated with other wavebands. 

Follow-up observations have further characterized the nature of some MSPs in the 4FGL-DR3 catalog \citep{Abdollahi2022}. For instance, 1SXPS J042749.2-670434 and 4FGL J1544.5-1126 have been verified as transitional MSP binaries \citep{Kennedy2020,Jaodand2021,Mata2023}, while 4FGL J0940.3-7610 has been identified as a candidate redback MSP binary through a follow-up observation \citep{Swihart2021}. Additionally, 4FGL J0336.0+7502 is classified as a black widow binary, and 4FGL J0212.1+5321, 4FGL J0523.3-2527, and 4FGL J0838.7-2827 have been confirmed as redback MSP candidates based on periodic optical and X-ray observations \citep{Halpern2017,Li2021,Halpern2022a}.  
It is noteworthy that 4FGL J0212.1+5321 and 4FGL J0838.7-2827 have recently been confirmed as radio MSPs \citep{Perez2023, Thongmeearkom2024}.
Despite the observational certifications of these sources, the 4FGL-DR3 catalog still comprises 2,157 sources unassociated to other wavelength bands. 

In this study, we performed a cross-matching analysis between 2,157 Fermi unassociated sources and the ZTF data to identify their corresponding optical counterparts. Taking into account the 4FGL-DR3 catalog's 4FGL 95 percent confidence error ellipses for gamma-ray sources, along with the foundation laid by \citet{Braglia2020}, we further defined our selection criteria as follows: 

(i) Choosing 4FGL-DR3 sources located within the ZTF survey sky, specifically those with declination greater than -28 degrees. 

(ii) Selecting ZTF targets within a circular region of 6 arcminutes radius centered on the 4FGL-DR3 gamma-ray sources.

While these selection criteria may not be commonly used in multi-wavelength identification of gamma-ray sources, they have been employed in order to maximize the coverage of the entire 4FGL error ellipses. After applying the aforementioned selection criteria, we successfully matched 1,351 gamma-ray sources to the ZTF data, accounting for approximately 62\% of the total number of Fermi unassociated sources.

\begin{figure*}[htpb]
\begin{center}
\includegraphics[width=1.0\textwidth]{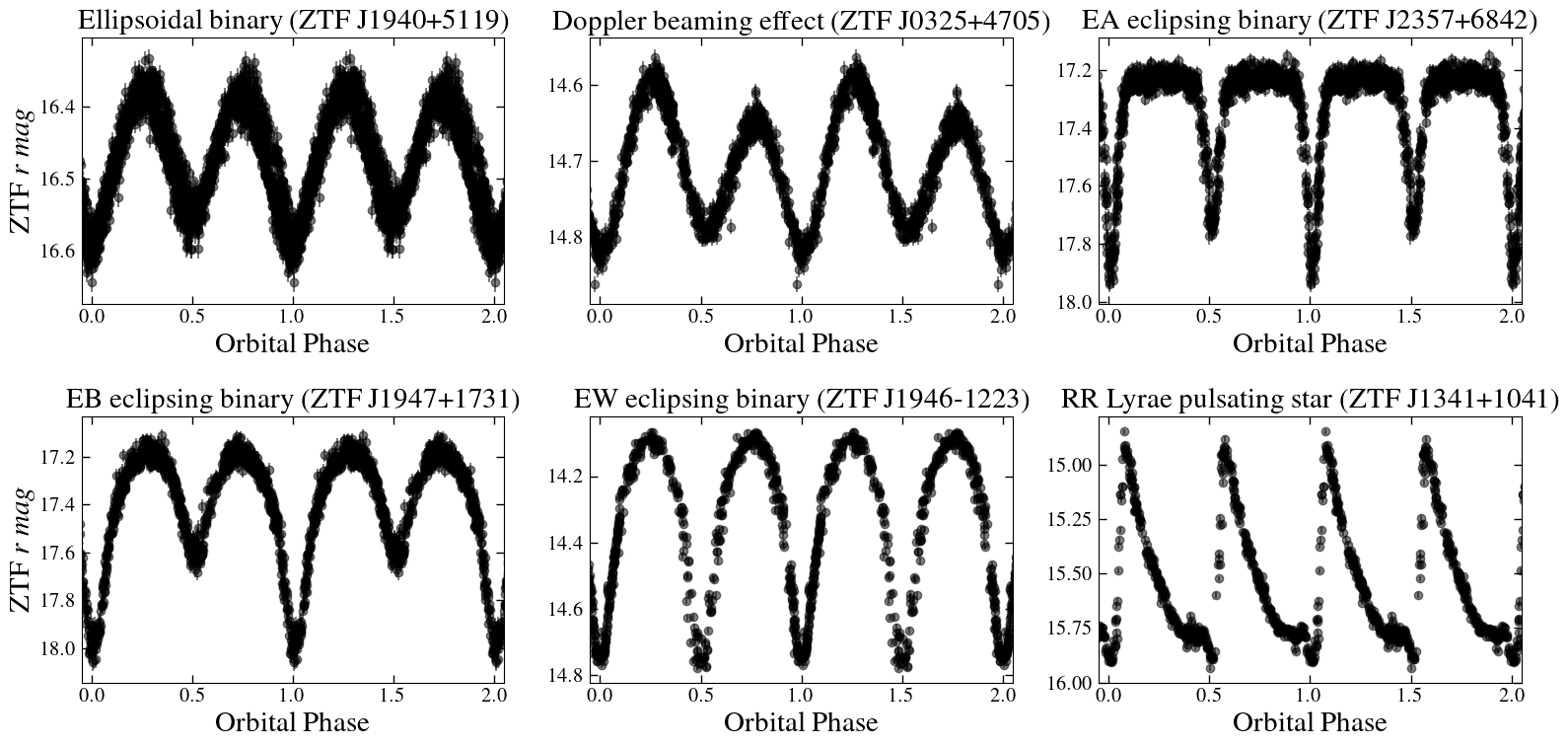}
\caption{Examples of various classes of variable stars identified near 4FGL-DR3 unassociated gamma-ray sources with optical counterparts from the ZTF data. Upper left: Ellipsoidal binaries, which exhibit quasi-sinusoidal variations with two distinct minima within a single orbital period, attributed to tidal deformation. Upper center: The Doppler beaming effect, a special type of ellipsoidal binary. Upper right: EA eclipsing binaries, characterized by light curves that remain almost constant or vary insignificantly between eclipses, with two minima per cycle or the absence of secondary minima. Bottom left: EB eclipsing binaries, which display two minima per cycle, and the secondary minimum depth is typically much smaller than the primary minimum. Bottom center: EW eclipsing binaries, composed of ellipsoidal components that are nearly in contact, with light curve minima depths that are nearly equal or differ only slightly. Bottom right: RR Lyrae type pulsating stars, with light curves that rapidly brighten and then slowly fade.}
\label{fig:examples}
\end{center}
\end{figure*}

\subsection{ZTF Photometry}

The ZTF is a 48-inch Schmidt telescope located at the Palomar Observatory, designed for conducting time-domain photometric surveys of the northern sky. The ZTF utilizes a large field of view of 47 $\rm deg^2$ and high-frequency sampling to capture the time variation of objects, with the sky area being covered approximately every 3 days. It offers unprecedented sensitivity and accuracy. The ZTF photometric data provides measurements of the brightness and variability of celestial sources in multiple filters (ZTF $g$-, $r$-, and $i$-bands), enabling the study of transients and variability across broad-bands. ZTF has a median $5 \sigma$ limiting magnitude of 20.8 in $g$-band, 20.6 in $r$-band, and 20.2 in $i$-band in a $30\rm\,s$ exposures.

\subsection{Light Curve Selection}

We will proceed with further exploration of two types of variable light curves, namely pulsar irradiation or ellipsoidal variations, within the test and search samples. To conduct large-scale processing of the variable light curve data and search for periodicity, it is necessary to establish a data processing pipeline. This pipeline consists of the following three components:

(i) Downloading the corresponding ZTF archived data within a 6-arcminute radius of each gamma-ray source from the ZTF Science Data System (ZSDS) and acquiring the ZTF-g and ZTF-r band light curves for the target sources.

(ii) Applying a period finding algorithm to individually process the light curve data of each ZTF source in order to search for periodic signals.

(iii) Generating folded light curves for each ZTF source.

The ZTF Science Data System (ZSDS) is a data system associated with the Zwicky Transient Facility (ZTF) project. It is designed to manage and provide access to the observational data collected by the ZTF survey. The ZSDS includes various components and functionalities for data storage, processing, and analysis. It allows researchers to retrieve and analyze ZTF data, including light curves and images, enabling scientific investigations and discoveries in the field of time-domain astronomy.

We first obtained the coordinates of each ZTF source within a 6-arcminute radius of the corresponding gamma-ray source.
Then, utilizing \textbf{wget}, we downloaded the corresponding light curve data from the ZSDS. We conducted a large-scale download of the light curve data for each target source based on ZTF DR10 from the NASA/IPAC Infrared Science Archive (IRSA)\footnote{\url{https://www.ipac.caltech.edu}}$^{,}$\footnote{\url{https://www.ztf.caltech.edu/page/dr8}}, successfully implementing this functionality using \textbf{wget}. \textbf{wget} is a command-line utility employed for retrieving files from web servers using the HTTP, HTTPS, and FTP protocols.

There is a wide range of algorithms available for identifying and extracting periodic signals from light curve data, and some of them provide open-source software packages. The main algorithms used for period finding search include the Lomb-Scargle Periodogram algorithm and its variants:

$\bullet$ \textbf{Multiband Lomb-Scargle Periodogram:} It incorporates common variability by regularizing the basic model, effectively simplifying the overall model and reducing background signals in the periodogram.

$\bullet$ \textbf{Trended Lomb-Scargle Periodogram:} It extends the basic Lomb-Scargle Periodogram algorithm by adding a linear trend parameter in time. This parameter can be useful when the underlying time series exhibits non-stationarity.

$\bullet$ \textbf{Super-Smoother:} This non-parametric adaptive smoothing algorithm applies a super-smooth technique to the time series data for each candidate frequency and calculates the model's error. The candidate frequency with the minimized model error represents the best-fit period. 

The \textbf{Gatspy} \footnote{\url{http://www.astroml.org/gatspy/}} website provides access to the open-source code for these algorithms.

In addition to the Lomb-Scargle algorithm, other period finding algorithms include Fast Template Periodogram, Conditional Entropy (CE) or Generalized Conditional Entropy (GCE), Box-least squares (BLS) algorithm, Analysis Of Variance (AOV) algorithm, and Phase Dispersion Minimization (PDM, PDM2) algorithm provided by \textbf{P4J} \footnote{\url{https://github.com/phuijse/P4J}}. These algorithms offer different techniques for identifying periodic signals in light curve data.

In large-scale survey data, it is often necessary to process light curve data for millions or even tens of millions of target sources. Choosing the appropriate period finding algorithm is crucial to save time and computational resources. Therefore, optimizing and accelerating the data processing pipeline with the right algorithm is of paramount importance.

We have constructed a data processing pipeline based on the Lomb-Scargle Periodogram algorithm. The LS algorithm is effective in identifying the periodicity of various types of variable sources, particularly sensitive to sources exhibiting sinusoidal signal characteristics. This sensitivity makes it particularly favorable for searching for ellipsoidal variations in spider pulsar binary systems.


To identify sources that are likely redbacks and black widows within our test and search samples, we have summarized the following criteria based on the test sample:

(i) Light curve exhibiting two types of characteristics:
The first type is ellipsoidal variations characterized by alternating primary maxima and minima. This ensures that the donor star dominates the light curve rather than an accretion disk. This feature has also been used by \citet{El-Badry2021} to search for evolved CVs and proto-ELM WDs.
The second type is pulsar irradiation, which shows a single peak within one orbital period ($\phi \sim 0.0 - 1.0$) and has extremely low brightness, averaging magnitudes beyond 20-21 mag.

(ii) For ellipsoidal variations\citep{El-Badry2021}, the peak-to-peak variability amplitude in the g/r band should be greater than 15\%. This criterion selects binary systems where the stars are tidally distorted and almost or completely Roche-lobe filling.

(iii) 
For pulsar irradiation, light curve of this type of binary system is sinusoidal, with a period equal to the orbital period, and maximum brightness coinciding with the passage of the pulsar in front of its companion, exhibiting characteristics of reflection effects. This criterion selects systems where one side of the star is heated by the pulsar.

\begin{figure*}[htpb]
\begin{center}
\includegraphics[width=1.0\textwidth]{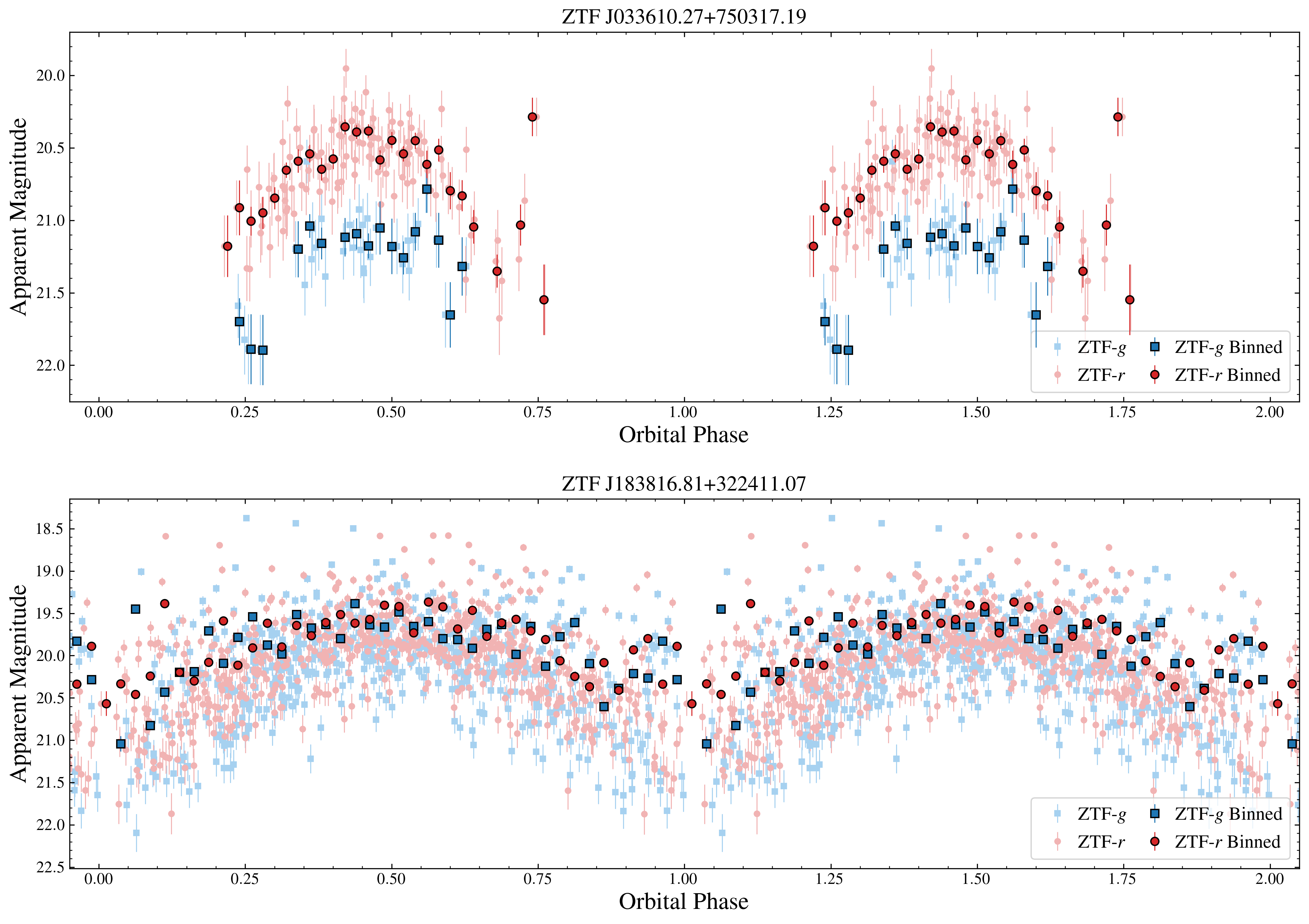}
\caption{The phase-folded light curves of the irradiation binary ZTF J0336+7502 and ZTF J1838+3224 exhibit the distinctive features of this class of variable stars. These characteristics are attributed to the heating of the companion star's side facing the pulsar, which results in a higher temperature compared to the opposite side. Notably, the light curve of ZTF J1838+3224 includes instances of flaring activity.}
\label{fig:IrradiationSample}
\end{center}
\end{figure*}

\subsection{Final sample}

Based on the sample selection and data processing pipeline described above, we obtained the light curves of ZTF targets within a 6 arcminute radius for both the test and search samples. Through visual inspection of the light curves, we found the following results:

For the test sample, out of a total of 42 known RB and BW systems, we identified 11 light curves exhibiting periodic variations, with 5 belonging to RB systems and 6 to BW systems. 
The figures \ref{fig:possimage1}-\ref{fig:possimage2} illustrate the POSS-1 red band survey images from the Virtual Observatory (VO) archives for 12 known spider pulsars within the test samples. The blue ellipses represent the 95\% confidence level 4FGL error ellipses, while the black circles indicate the search areas covered by ZTF data. It is observed that the optical counterparts of the known spider pulsars are in close proximity to the positions of the 4FGL gamma-ray sources, with both being contained within the Fermi error ellipses. The 4FGL-DR3 data provides a high degree of precision in the localization of gamma-ray sources for the known spider pulsars, facilitating the identification of ZTF optical counterparts in the vicinity of the gamma-ray source coordinates. This offers a foundation for the subsequent search for spider pulsar candidates in the vicinity of Fermi unassociated gamma-ray sources.
Figure \ref{fig:testsample} shows the 12 spider pulsar binary systems with periodic variations identified in the test sample (see Table \ref{tab:knowspiders}), including 4 light curves exhibiting ellipsoidal variations and 8 light curves displaying pulsar irradiation features. Table \ref{tab:knowspiders} shows the coordinates, orbital periods, and photometric variability types of the 12 known Spider pulsar binary systems. Figures \ref{fig:LSPSD1} and \ref{fig:LSPSD2} display the frequency spectra derived from the Lomb-Scargle periodograms of the known spider pulsars in our test sample, alongside the frequencies corresponding to the best-fit periods.

For the search sample, a total of 1,351 Fermi unassociated sources ($\rm decl. > -28^{\circ}$) were matched to over 20 million ZTF targets. After analysis through the data processing pipeline, 1,566 sources were identified to have periodic variations in their light curves. We consider the classification scheme provided by \citet{Ren2023}, and further categorize the light curves into the following types: EA-type binary stars, EB-type binary stars, EW-type binary stars, and ELL-type binary stars, as well as other unknown types of light curve shapes.

EA-type eclipsing binary systems, synonymous with $\beta$ Persei-type (Algol) systems, consist of close binaries with stellar components that are spherical or slightly ellipsoidal and do not engage in physical contact, generally constituting a detached configuration. A subset of these systems is categorized as close semi-detached binaries, arising from one component star having expanded to fill its Roche lobe. The light curve characteristics of these systems exhibit minimal variation in luminosity during the eclipse phase, which can be attributed to reflection effects, minor ellipsoidal distortions of the binary components, or intrinsic physical variations. The upper right panel of Figure \ref{fig:examples} displays an example of the light curve shape.

EB-type binary stars, also known as $\beta$ Lyrae-type eclipsing systems, are close binary systems composed of ellipsoidal stars. The two stellar components are in such close proximity that they gravitationally distort each other, yet they do not make physical contact, thus precluding mass exchange. The light curve characteristics of these systems are such that there is no flat, well-defined phase of constant brightness between eclipses, making it impossible to specify the exact times of the onset and end of eclipses. The bottom left panel of Figure \ref{fig:examples} shows an example of the light curve shape. 

EW-type binary stars, also known as W Ursae Majoris-type eclipsing variables, are close binary systems composed of two nearly touching ellipsoidal FGK dwarfs. Both constituent stars typically fill their critical Roche lobes, have very similar masses, and share a common envelope, leading to significant interactions including mass transfer. The light curve characteristics of these systems include nearly equal depths for the primary and secondary minima, with continuous light variations, making it impossible to precisely specify the start and end times of the eclipses. The bottom center panel of Figure \ref{fig:examples} displays an example of the light curve shape.

ELL-type binary stars, or ellipsoidal variables (EV), constitute a category of close binary systems with ellipsoidal components that do not undergo eclipses. The characteristic light curves of these systems display primary and secondary minima of differing depths. The underlying photometric variability is attributed to two principal effects: the alteration of the stellar cross-section as perceived by the observer, arising from the ellipsoidal distortion due to tidal interactions, and the reflection effect, which involves the absorption and re-emission of radiation by each component emanating from its companion. The upper left (and upper center) panel of Figure \ref{fig:examples} illustrates an example of the light curve shape.

Based on the previous description and the classification criteria provided by \citet{Ren2023} and \citet{Skarka2022}, these sources can be further categorized as follows: 79 EA-type binary stars, 24 EB-type binary stars, 194 EV-type binary stars (ELL-type binary stars), 131 EW-type binary stars, 75 RR Lyrae type pulsating stars, 1,063 variable stars of unknown type. Figure \ref{fig:examples} shows several representative examples of light curves from periodic variable sources.

\begin{figure*}[htpb]
\begin{center}
\includegraphics[width=1.0\textwidth]{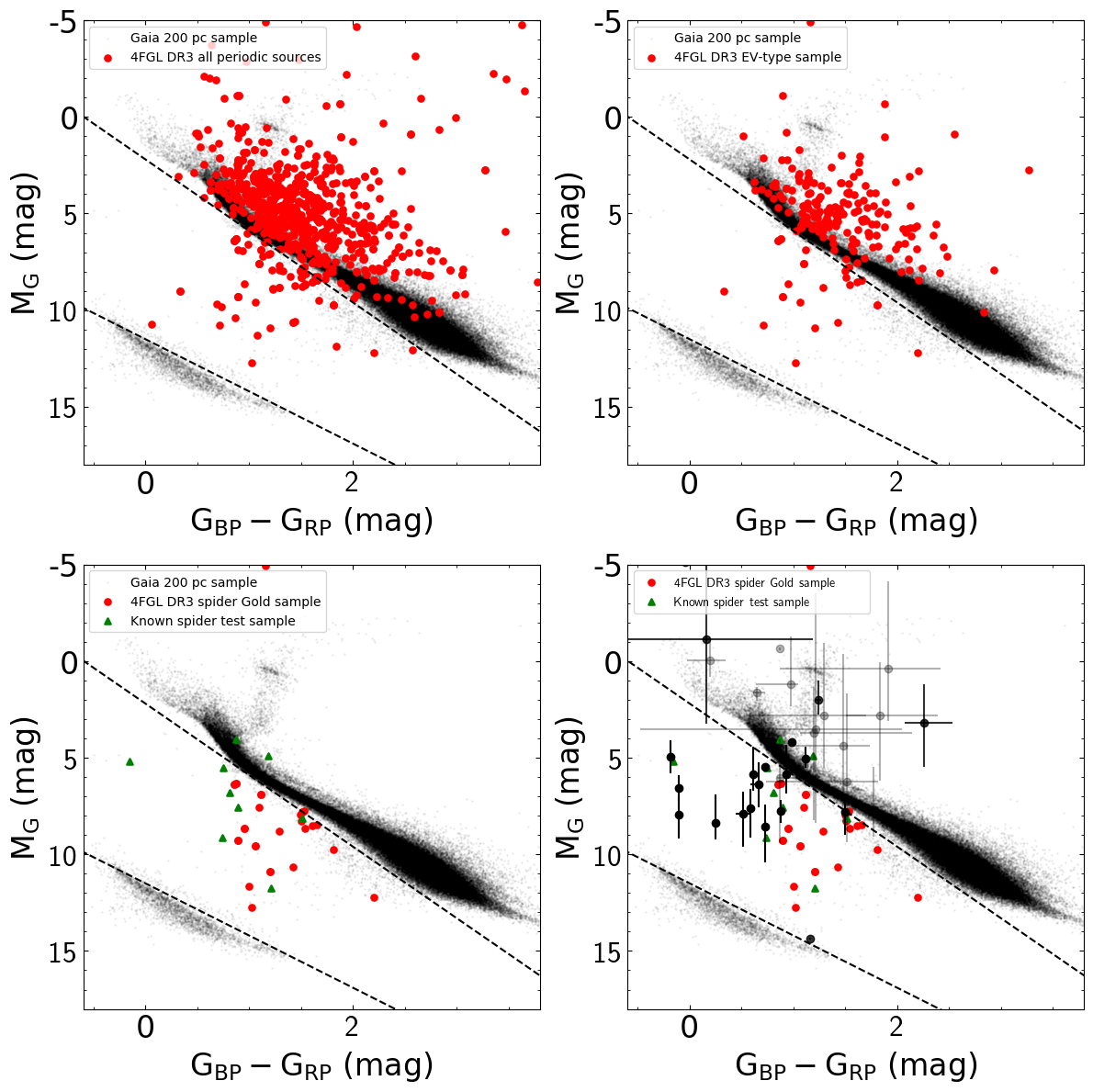}
\caption{The location of the sample selection on the Gaia color-magnitude diagram. In all panels, the scatter points and the overlaid two-dimensional histogram represent the stellar density of the Gaia 200 parsec background sources. The black dashed line indicates the selection boundary for the golden sample of millisecond pulsars using the Gaia Hertzsprung-Russell diagram. Upper left: The red points denote all ZTF optical counterparts exhibiting periodic variability found within a 6 arcminute range of the 4FGL-DR3 unassociated gamma-ray sources. Upper right: The ellipsoidal variable binaries, also known as EV-type binaries, among the ZTF optical counterparts. Bottom left: The red pentagons represent the 24 golden sample stars, while the green triangles represent the 12 test sample stars. Bottom right: The red pentagons represent the 24 golden sample stars, the green triangles represent the 12 test sample stars, the black circles indicate confirmed millisecond pulsar companions, and the grey circles are candidate MSP companions \citep{Antoniadis2021,Swihart2022a}.}
\label{fig:gaiahr}
\end{center}
\end{figure*}  

\begin{figure*}[htpb]
\begin{center}
\includegraphics[width=1.0\textwidth]{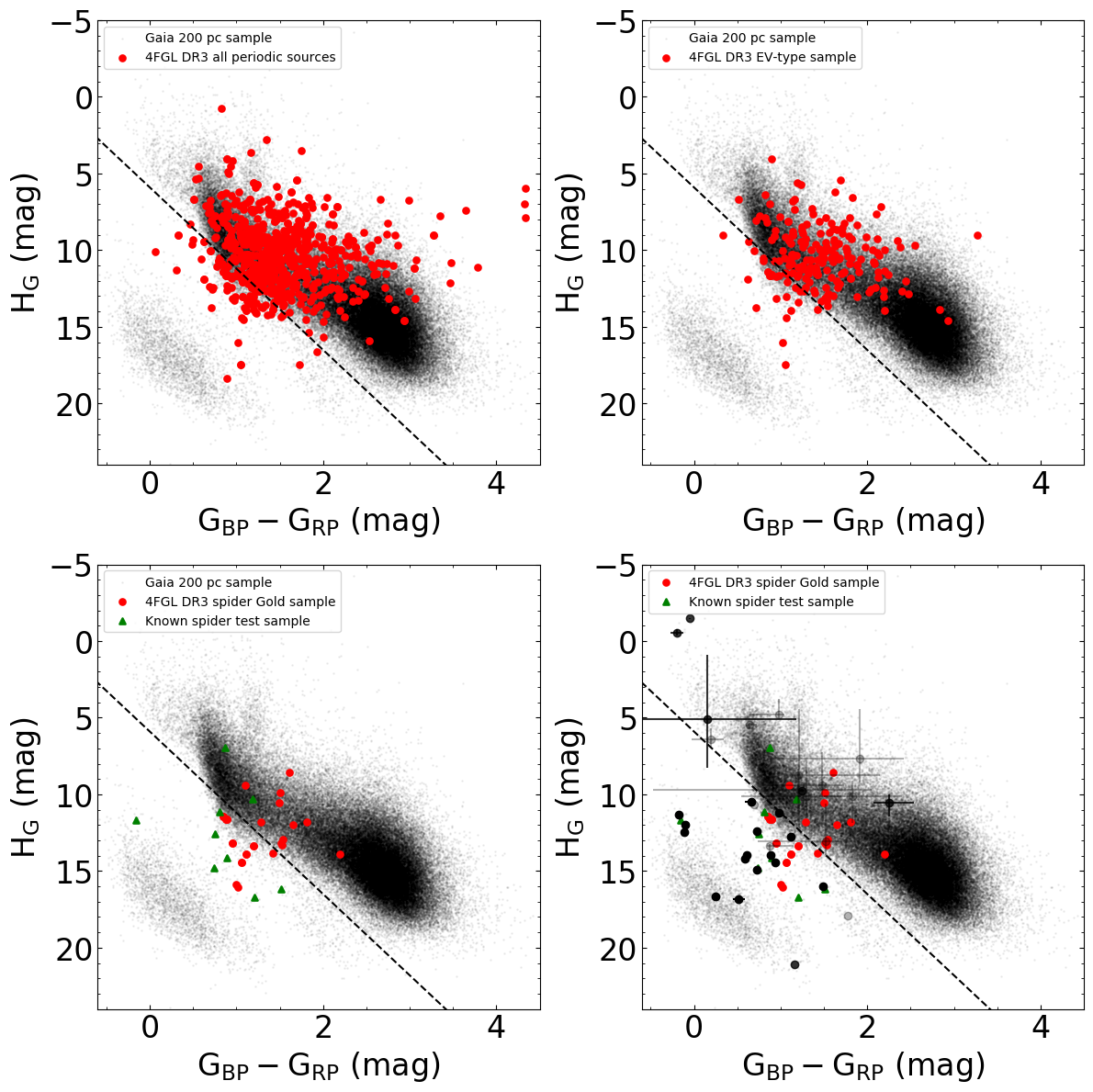}
\caption{The location of the sample selection on the Gaia color-reduced proper motion (RPM) diagrams. In all panels, scatter points and the overlaid two-dimensional histogram illustrate the stellar density of the Gaia 200 parsec background sources. The black dashed line represents the selection boundary for the golden sample of millisecond pulsars, utilizing the Color-RPM diagram as proposed by \citet{Antoniadis2021}. Upper left: The red points indicate all ZTF optical counterparts exhibiting periodic variability found within a 6 arcminute range of the 4FGL-DR3 unassociated gamma-ray sources. Upper right: The ellipsoidal variable binaries, also known as EV-type binaries, among the ZTF optical counterparts. Bottom left: The red pentagons denote the 24 stars of the golden sample, while the green triangles represent the 12 stars of the test sample. Bottom right: The red pentagons denote the 24 stars of the golden sample, the green triangles represent the 12 stars of the test sample, the black circles indicate confirmed millisecond pulsar companions, and the grey circles represent candidate MSP companions \citep{Antoniadis2021,Swihart2022a}.}
\label{fig:gaiarpm}
\end{center}
\end{figure*}

\section{Analysis and Results} \label{sec:analaysisresults}

\subsection{Irradiation Binary Sample}

Based on the analysis of light curve shapes from the search sample's ZTF sources, we identified one source, ZTFJ033610.27+750317.19 (hereafter ZTF J0336+7502), exhibiting light curve characteristics consistent with pulsar irradiation features. ZTF J0336+7502 has an orbital period of 223.09 minutes (3.718167 hours), and coordinates of $\rm \alpha (J2000) = 03^{h}36^{m}10^{s}.27, \delta (J2000) = +75^{d}03^{m}17^{s}.19$, a Gaia G-band average magnitude of 20.612339 mag, and a color index $\rm G_{BP}-G_{RP}$ = 1.0511 mag. Due to its magnitude reaching the limiting magnitude of Gaia photometry, the provided negative parallax ($\varpi$) from Gaia DR3 is -0.8516 mas. This source lies in the 95\% error region of the gamma-ray source 4FGL J0336.0+7502 and displays various pulsar-like evidence, recorded for the first time in the 4FGL-DR2/DR3 catalog. 

\citet{Li2021} conducted optical photometric observations of this source using telescopes and also found an X-ray counterpart near 4FGL J0336.0+7502. Combining optical, X-ray, and gamma-ray observations confirmed that this is a black widow binary system. Li et al. provided a period of $\rm P_{orb}$ = 3.718178(9) hours, which closely matches our calculated result, further validating the use of ZTF data in the search for pulsar binary systems. 
The upper panel of Figure \ref{fig:IrradiationSample} shows the folded light curves of ZTF J0336+7502 in the ZTF g- and r-bands.

In our search sample, an additional Irradiation-type binary system with an orbital period of 241.493 minutes was identified: ZTF J183816.81+322411.07 (hereafter ZTF J1838+3224). The optical counterpart coordinates of ZTF J1838+3224 in the ZTF are $\rm \alpha (J2000) = 18^{h}38^{m}16^{s}.81, \delta (J2000) = +32^{d}24^{m}11^{s}.07$. The Gaia G-band average magnitude is 20.60957 mag, with a color index of $\rm G_{BP}-G_{RP} = 0.9992714$ mag. The Gaia parallax is 1.6056 mas, which corresponds to an estimated distance of 3.41550 kpc based on the Gaia parallax parameters. This candidate is located within the 95\% elliptical error contour of the Fermi Gamma-ray source 4FGL J1838.2+3223.

Recently, \citet{Zyuzin2024} also utilized ZTF data to discover an optical counterpart near 4FGL J1838.2+3223, with an orbital period of 4.02 hours. They conducted follow-up observations based on optical time-series multi-band photometry and authenticated that 4FGL J1838.2+3223 is a flaring ‘spider’ pulsar candidate. Through light curve modeling, they estimated the companion mass to be 0.10±0.05 solar masses, with the side facing the pulsar intensely heated to approximately 11300±400 K, while the temperature on the opposite side is only 2300±700 K, and it does not fill its Roche lobe. The optical counterpart identified in our search sample corresponds to the same source reported in that study. The bottom panel of Figure \ref{fig:IrradiationSample} shows the folded light curve of ZTF J1838+3224, with ZTF data indicating the presence of flaring activity.

\subsection{Ellipsoidal Binary}

EV-type binary systems, also known as ellipsoidal variables, are a type of binary star system characterized by non-eclipsing reflection effects, exhibiting quasi-sinusoidal variations with two distinct minima within one orbital period. This quasi-sinusoidal variation in the light curve is caused by tidal deformation. 
However, there can be reflection effects in ellipsoidal variables, which can produce the effect of quasi-sinusoidal variations with two minima and two maxima, particularly observed in the light curves of some redback systems.
Such light curve characteristics have been observed in a variety of binary systems, including those composed of a hot white dwarf and a low-mass M dwarf \citep{Burdge2020,Ren2023}, evolved cataclysmic variables and proto-extremely low-mass white dwarfs (proto-ELM WDs) \citep{El-Badry2021}, and spider-pulsar binary systems \citep{Braglia2020}.
In EV-type binary systems, mass transfer between the accretor and the donor star usually occurs through stable Roche lobe overflow, indicating that the donor star has evolved to a late stage and is gradually forming a compact object. Therefore, the light curve features of ellipsoidal variables serve as an important criterion in the search for ultra-compact binary systems.

In our test sample, we have identified a total of four sources with light curves exhibiting ellipsoidal variations, and all of these sources belong to redback systems (see Table \ref{tab:knowspiders}). Redback systems are composed of a pulsar and a low-mass non-degenerate star, with the material of the companion star filling its Roche lobe before evolving into a compact object. Among the redback systems discovered so far, the majority of their light curves have been classified as EV-type.
It is worth noting that in the case of PSR J1048+2339, observations have shown that its light curve shape transitions between ellipsoidal variations and pulsar irradiation. During the ZTF observation period, the data obtained indicated pulsar irradiation for this system.

In our search sample, a total of 194 ZTF objects have been identified as exhibiting ellipsoidal light curve variability. Given that these objects were found in close proximity to Fermi-LAT gamma-ray sources, they are highly likely to be potential optical counterparts of the gamma-ray sources. Among these EV-type binary systems, there is a significant probability of discovering potential spider pulsar binary candidates, with a particular focus on redback systems. 

The upper-right panel of Figure \ref{fig:gaiahr} shows the distribution of the 194 candidate EV-type binary stars discovered in the Gaia Hertzsprung-Russell diagram from our search sample. Among these candidates, 22 stars are found to be situated in the region between the main sequence and the white dwarf cooling sequence, while the remaining 172 stars are either located within the main sequence or above it.

\subsection{The Gold Sample of Spider Pulsar Binary}
\subsubsection{Gaia Hertzsprung-Russell Diagram Selection}
In order to search for candidate spider pulsar binaries from the EV-type binary star sample, we further narrowed down the number of candidates based on their distribution in the Gaia Hertzsprung-Russell diagram. Following the systematic study by \citet{Antoniadis2021}, on how to discover pulsar companions in Gaia DR2 data, they cross-matched the known 1534 rotation-powered pulsars with Gaia DR2 data and searched for Gaia DR2 objects within a 20 arcsec radius. They identified 20 previously known binary companions, most of which belong to millisecond pulsar systems, and also discovered 8 new objects associated with young pulsars. 
The Gaia pulsar objects discovered by \citet{Antoniadis2021} primarily consist of companions to millisecond pulsars (MSPs). Among them, the majority of eclipsing/ellipsoidal MSPs are distributed in the region between the main sequence and the white dwarf cooling sequence in the Gaia Hertzsprung-Russell diagram. Utilizing this finding, we can effectively reduce the number of EV-type binary star samples, enabling an efficient search for redback and black widow candidates.

\citet{Antoniadis2021} discovered that three millisecond pulsars with low-mass white dwarf (WD) companions, along with the majority of eclipsing/ellipsoidal MSPs, are located below the main sequence in the Gaia Hertzsprung-Russell diagram. In this study, we adopt the search results from \citet{Antoniadis2021} and further define the selection criteria for the golden sample of spider pulsar binary systems, which satisfy the following conditions:
\begin{equation}\label{eq:1}
\begin{array}{rcl}
M_{\rm G}  & \leq   & 2.7(G_{\rm BP}-G_{\rm RP}) + 10.5,  \\
M_{\rm G}  & >  & 3.7(G_{\rm BP}-G_{\rm RP}) + 2.4, \\ 
\end{array}
\end{equation}
where $G_{\rm BP}$ and $G_{\rm RP}$ represent the mean magnitudes in the Gaia BP and RP photometric bands, respectively. These criteria serve as powerful tools for characterizing distinct kinematic populations without relying on precise parallax measurements. 

Based on the Gaia Hertzsprung-Russell diagram selection criteria mentioned above, we have identified a total of 22 EV-type binary stars that meet the requirements. These binaries are referred to as the golden sample of spider pulsar binary systems (see Table \ref{tab:goldspiders}), as shown in the lower-left panel of Figure \ref{fig:gaiahr}. The lower-right panel of Figure \ref{fig:gaiahr} illustrates the positions of the known spider pulsar binaries from the test sample and the newly discovered golden sample in the Hertzsprung-Russell diagram. The black dots represent the known pulsars and candidate associations matched by \citet{Antoniadis2021} et al. in Gaia DR2. We observe that a significant number of golden sample stars appear to have redder colors compared to the known pulsars. This discrepancy could possibly be attributed to the presence of more cataclysmic variables in the golden sample.

\subsubsection{Color-Reduced Proper Motion Diagram Selection}

\citet{Antoniadis2021} pointed out that in the absence of precise parallax measurements, particularly when affected by interstellar reddening \cite{green2014,green2018,green2019}, one can employ a quantity called the reduced proper motion, denoted as $H_{\rm g}$, to distinguish different kinematic populations. The expression for $H_{\rm g}$ is given by 
\begin{equation}\label{eq:2}
    H_{\rm g} = m_{\rm g} + 5\log_{10}{\frac{\mu}{{\rm mas\,yr^{-1}}}}-10+A_{\rm g}  
\end{equation}
where $m_{\rm g}$ and $A_{\rm g}(d)$ represent the Gaia band apparent magnitude and extinction, respectively. The term $\mu = \sqrt{\mu_{\alpha}^2 + \mu_{\delta}^2}$ corresponds to the magnitude of proper motion. The quantity $H_{\rm g}\equiv M_{\rm g} + 5\log_{10}(v_{\rm tr} /4.74057,{\rm km},{\rm s}^{-1})$ is analogous to the absolute magnitude but sensitive to the source's transverse velocity \citep{Antoniadis2021}. This makes it a valuable metric for distinguishing between different kinematic populations, even in the absence of a precise parallax measurement. 

Figure \ref{fig:gaiarpm} illustrates the distribution features of all periodic sources, EV-type binary stars, and the golden sample of spider pulsar binary systems in the color-$H_{\rm g}$ (referred to as RPM) diagrams from our search sample. Building upon the search results of \citet{Antoniadis2021}, we further define the constraints on $H_{\rm g}$ as follows:
\begin{equation}\label{eq:3}
\begin{aligned}
H_{\rm g} & > 5.3(G_{\rm BP}-G_{\rm RP}) + 5.9,
\end{aligned}
\end{equation}
Based on the further constraints on $H_{\rm g}$ in the RPM diagrams, the number of golden sample stars is reduced to nine, as shown in the lower-right panel of Figure \ref{fig:gaiarpm}.

The application of Equations (\ref{eq:1}) and (\ref{eq:3}) proves to be highly effective in filtering out field stars. However, some overlap is expected with hot subdwarfs and extremely low-mass white dwarfs and their progenitors \citep{Pelisoli2019,El-Badry2021,Ren2023}. \citet{Antoniadis2021} proposed that this overlap can be mitigated by employing additional criteria, such as $H_{\rm g} \leq 6.9(G_{\rm BP} - G_{\rm RP}) + 13.0$, to filter out white dwarfs. Alternatively, the use of multiwavelength observations (e.g., radio, X-ray, gamma-ray counterparts) can be employed for confirmation purposes.

\subsubsection{4FGL 95\% Confidence Error Ellipse Selection}
In the study by \citet{Braglia2020}, public repositories of optical and X-ray data were employed to cross-match with unidentified gamma-ray sources listed in the 3FGL, with the aim of discovering black widow and redback analogs, culminating in the identification of two prospective objects. The investigative protocol was constructed to detect counterparts in optical and X-ray wavelengths that reside within the 95\% confidence error ellipse of the 4FGL gamma-ray sources. An expansion of this search within the ZTF dataset led to the precise localization of 24 golden sample candidates in the immediate 6-arcminute perimeter surrounding the gamma-ray sources. Upon a more detailed examination of the candidates confined within the 4FGL's 95\% confidence error ellipse boundaries, it was determined that 19 of the 24 spider millisecond pulsar gold sample candidates met the established criteria. Five candidates were discerned to be situated beyond the 95\% confidence error ellipse of the 4FGL: ZTFJ1741-1620, ZTFJ1820-1524, ZTFJ1957+1233, ZTF J2144+7714, and ZTFJ1816+1747, as delineated in Table \ref{tab:Golddistance}. Appendix B presents the cross-match outcomes with respect to the 4FGL's 95\% confidence error ellipse and the ZTF's 6-arcminute boundary for the 24 gold sample candidates.

\begin{figure*}[htpb]
\begin{center}
\includegraphics[width=1.0\textwidth]{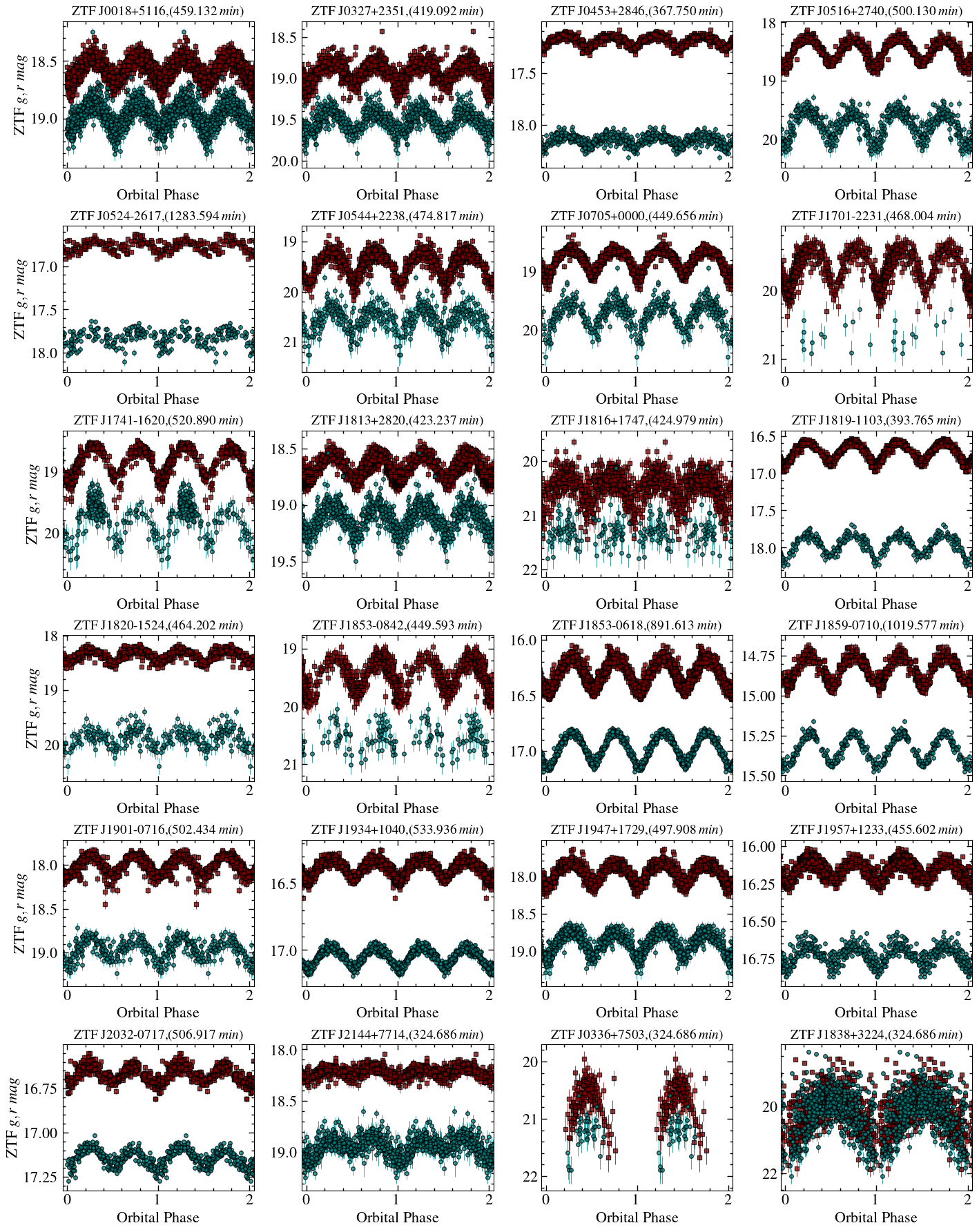}
\caption{Phase-folded ZTF light curves of the 24 millisecond pulsar Gold sample selected using the Gaia H-R diagram.}
\label{fig:goldsample}
\end{center}
\end{figure*}

\begin{deluxetable*}{ccccccccc}
\tablenum{2}
\tablecaption{Coordinates and Gaia parameters of the 24 Spider Gold sample candidates. \label{tab:goldspiders}}
\tablewidth{0pt}
\tablehead{
\colhead{4FGL Name} & \colhead{ZTF Name} & \colhead{Gaia DR3} &  \colhead{RA}  & \colhead{Dec} & \colhead{$\varpi$} & \colhead{G} & \colhead{$G_\mathrm{BP}-G_\mathrm{RP}$} &
\colhead{$P_\mathrm{orb}$}  \\
\colhead{} & \colhead{} & \colhead{sourceID} & \colhead{(deg)} & \colhead{(deg)} & \colhead{(mas)} & \colhead{(mag)} &  \colhead{(mag)} & \colhead{(min)} 
}
\startdata
4FGL J0018.8+5112 & ZTF J0018+5116 & 394965712236923392  & 4.613016   & 51.271605 & 0.3312 & 18.60613 & 0.852434 & 459.132 \\
4FGL J0327.3+2355c & ZTF J0327+2351 & 68712810947091072   & 51.929282  & 23.862197 & 0.5137 & 18.917133 & 1.099457 & 419.092 \\
4FGL J0453.3+2843 & ZTF J0453+2846 & 155048662983325184  & 73.291557 & 28.772751 & 1.4368 & 17.46426 & 1.506676 & 367.75 \\
4FGL J0516.7+2743 & ZTF J0516+2740 & 3421457779348865152 & 79.208417  & 27.672955 & 0.9365 & 18.522814 & 1.653307 & 500.13 \\
4FGL J0524.5-2614 & ZTF J0524-2617 & 2956906454393311744 & 81.143796  & -26.290986 & 1.4935 & 16.848598 & 1.530293 & 887.082 \\
4FGL J0544.4+2238 & ZTF J0544+2238 & 3403510274955026944 & 86.074310 & 22.643846 & 0.632 & 19.427324 & 1.609268 & 474.817 \\
4FGL J0705.8-0004 & ZTF J0705+0000 & 3112746864531854336 & 106.470333 & 0.006606 & 0.6469 & 18.802464 & 1.494776 & 449.656 \\
4FGL J1701.8-2226 & ZTF J1701-2231 & 4113999849726115072 & 255.447897 & -22.523927 & 1.0417 & 19.60232 & 1.812256 & 468.004 \\
4FGL J1741.1-1617 & ZTF J1741-1620 & 4124121644558712960 & 265.376582 & -16.337408 & 0.6103 & 18.797035 & 1.515532 & 520.89 \\
4FGL J1813.5+2819 & ZTF J1813+2820 & 4589333153195186432 & 273.359481 & 28.335530 & 0.3071 & 18.685806 & 0.877638 & 423.237 \\
4FGL J1816.7+1749 & ZTF J1816+1747 & 4523274116642584064 & 274.116961 & 17.786791 & 1.033 & 20.522554 & 1.425505 & 370.204 \\
4FGL J1819.4-1102 & ZTF J1819-1103 & 4154315810078443392 & 274.952160 & -11.061632 & 2.0133 & 20.663109 & 2.197966 & 393.765 \\
4FGL J1819.9-1530 & ZTF J1820-1524 & 4098113830845484928 & 275.013310 & -15.406315 & 1.1123 & 20.492567 & 0.710396 & 464.194 \\
4FGL J1853.2-0841 & ZTF J1853-0842 & 4203849118139045760 & 283.321187 & -8.71233 & 0.7924 & 19.096235 & 1.539402 & 449.593 \\
4FGL J1853.6-0620 & ZTF J1853-0618 & 4253939878633735296 & 283.348575 & -6.311358 & 0.5217 & 19.940546 & 0.950073 & 891.613 \\
4FGL J1859.2-0706 & ZTF J1859-0710 & 4205576146028340864 & 284.827387 & -7.166973 & -0.0274 & 19.074621 & 1.160341 & 1019.577 \\
4FGL J1901.8-0718 & ZTF J1901-0716 & 4205374935376962304 & 285.483774 & -7.280824 & 1.5958 & 19.858643 & 1.202694 & 502.434 \\
4FGL J1934.2+1036 & ZTF J1934+1040 & 4314629729310701312 & 293.513968 & 10.667270 & 2.6343 & 20.604877 & 1.020214 & 533.936 \\
4FGL J1947.1+1729 & ZTF J1947+1729 & 1821056517089353216 & 296.833042 & 17.490447 & 1.2935 & 18.216074 & 1.286856 & 497.908 \\
4FGL J1957.6+1230 & ZTF J1957+1233 & 4304207561657130752 & 299.448154 & 12.555031 & 0.9454 & 19.637606 & 1.060019 & 455.602 \\
4FGL J2032.3-0720 & ZTF J2032-0717 & 6907532096123807104 & 308.034276 & -7.287635 & 2.9681 & 16.914207 & 0.89426 & 506.917 \\
4FGL J2144.2+7709 & ZTF J2144+7714 & 2283760245882786560 & 326.211604 & 77.243606 & 15.65 & 18.33 & 0.81 & 324.686 \\
\hline
4FGL J0336.0+7502$^{a}$ & ZTF J0336+7503 & 544927450310303104 & 54.042790 & 75.054776 & 0.8516 & 20.615477 & 1.0510998 & 223.091 \\
4FGL J1838.2+3223$^{b}$ & ZTF J1838+3224 & 2090923983890463104 &  279.570111 & 32.4031872 & 1.6056 & 20.60957 & 0.9992714 & 241.493 \\
\enddata
\tablecomments{The coordinates and basic photometric characteristics of the 24 Spider Gold sample candidates discovered so far using ZTF data. Coordinates are taken from \emph{Gaia} and are in J2000.0. The IAU name is provided in the catalog.
\tablerefs{(a) \citet{Li2021}, (b) \citet{Zyuzin2024}}}
\end{deluxetable*}

\begin{figure*}[htpb]
\begin{center}
\includegraphics[width=1.0\textwidth]{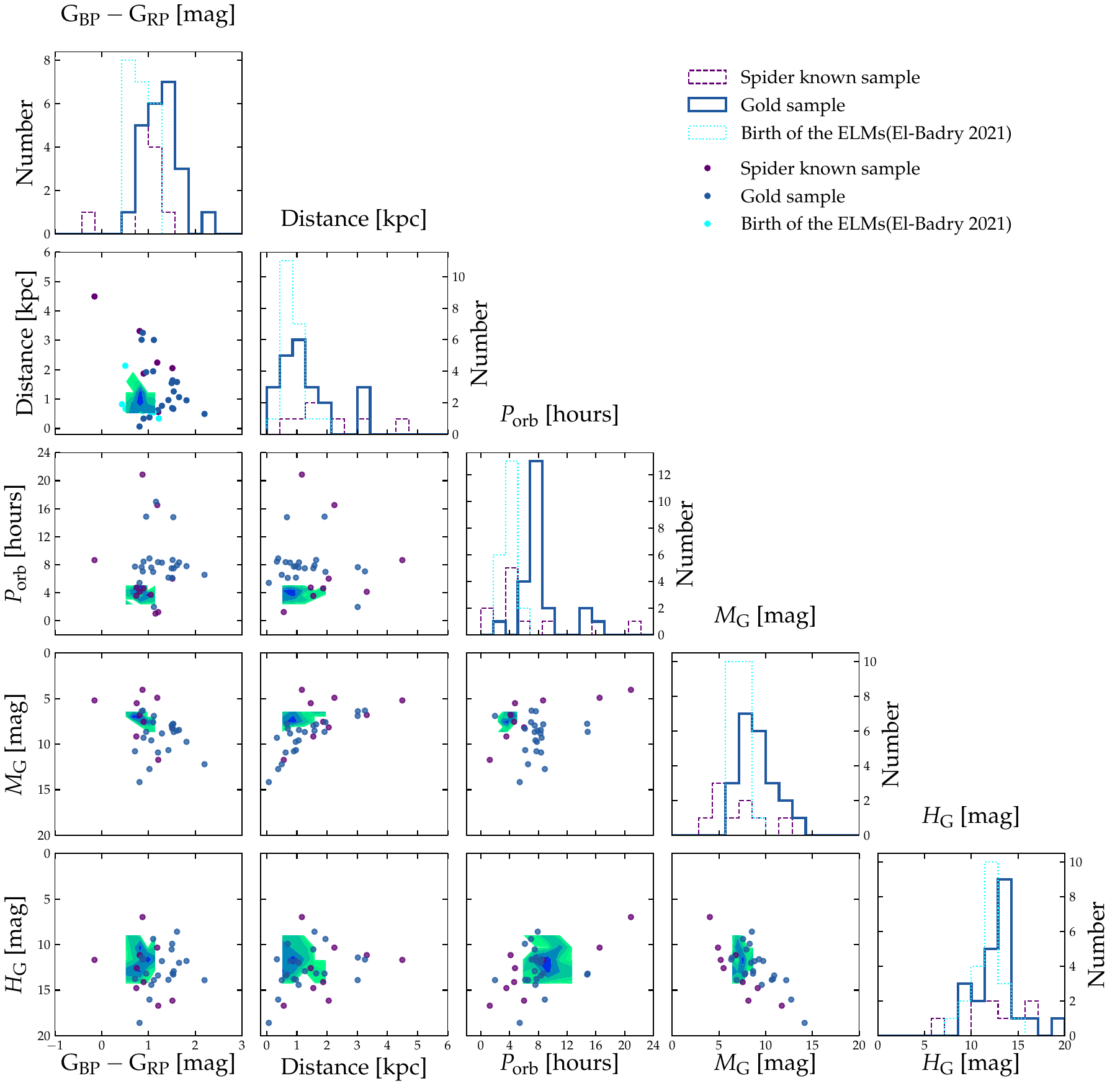}
\caption{The corner plot for the spider known sample, the Gold sample, and the 'Birth of the ELMs' as described by \citet{El-Badry2021} are provided. These plots depict the correlation distributions and one-dimensional histograms for various parameters, including the Gaia BP-RP color, distance, orbital period, Gaia G-band apparent magnitude, and the Gaia reduced proper motion.}
\label{fig:cornerplot}
\end{center}
\end{figure*}

\begin{figure*}[htpb]
\begin{center}
\includegraphics[width=1.0\textwidth]{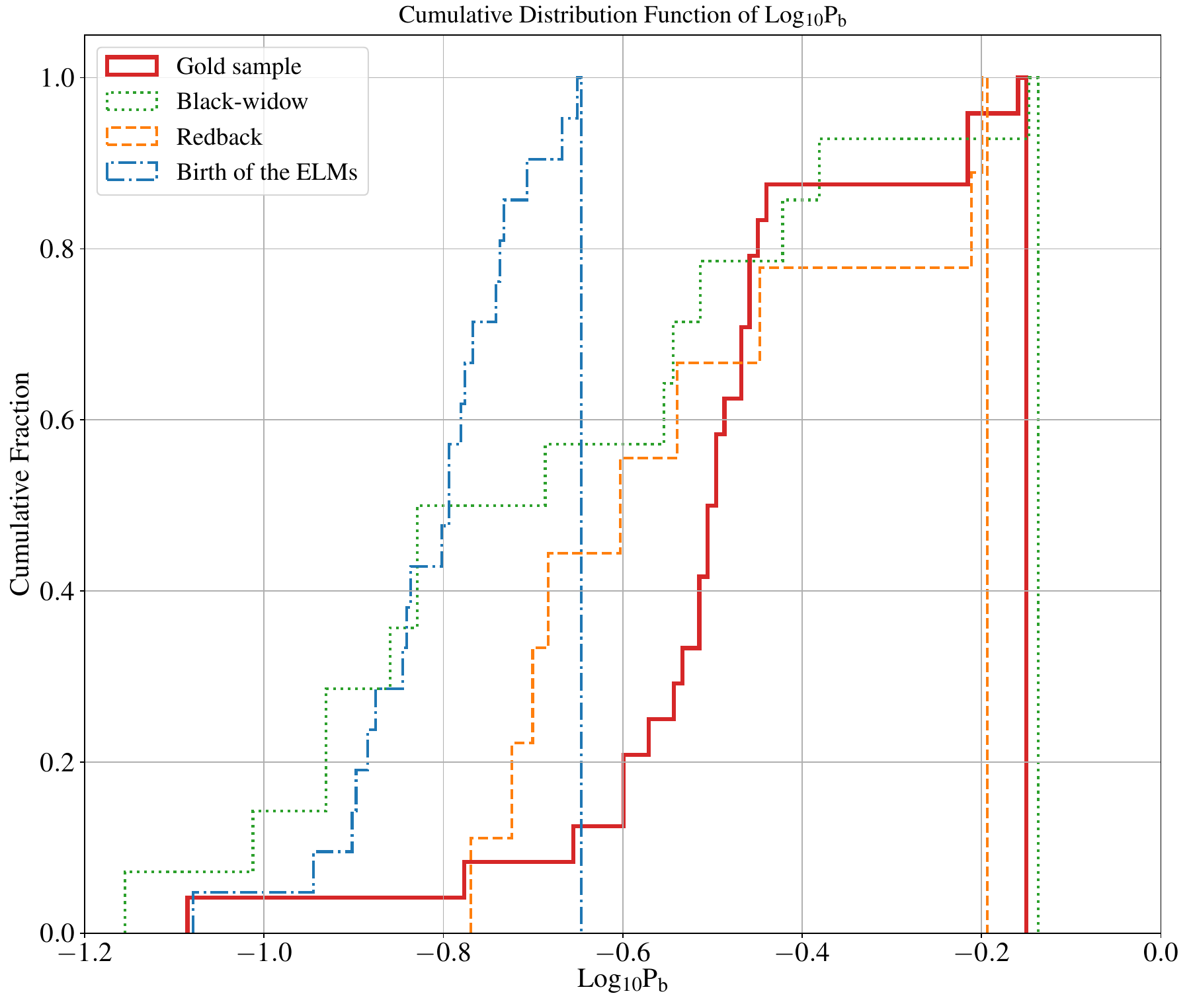}
\caption{Cumulative distribution functions of orbital period are presented for the known redback sample, the known black widow sample, the spider Gold sample candidates, and the 'Birth of the ELMs' sample, as described by \citet{El-Badry2021}.}
\label{fig:cdf}
\end{center}
\end{figure*}  

\begin{figure*}[htpb]
\begin{center}
\includegraphics[width=1.0\textwidth]{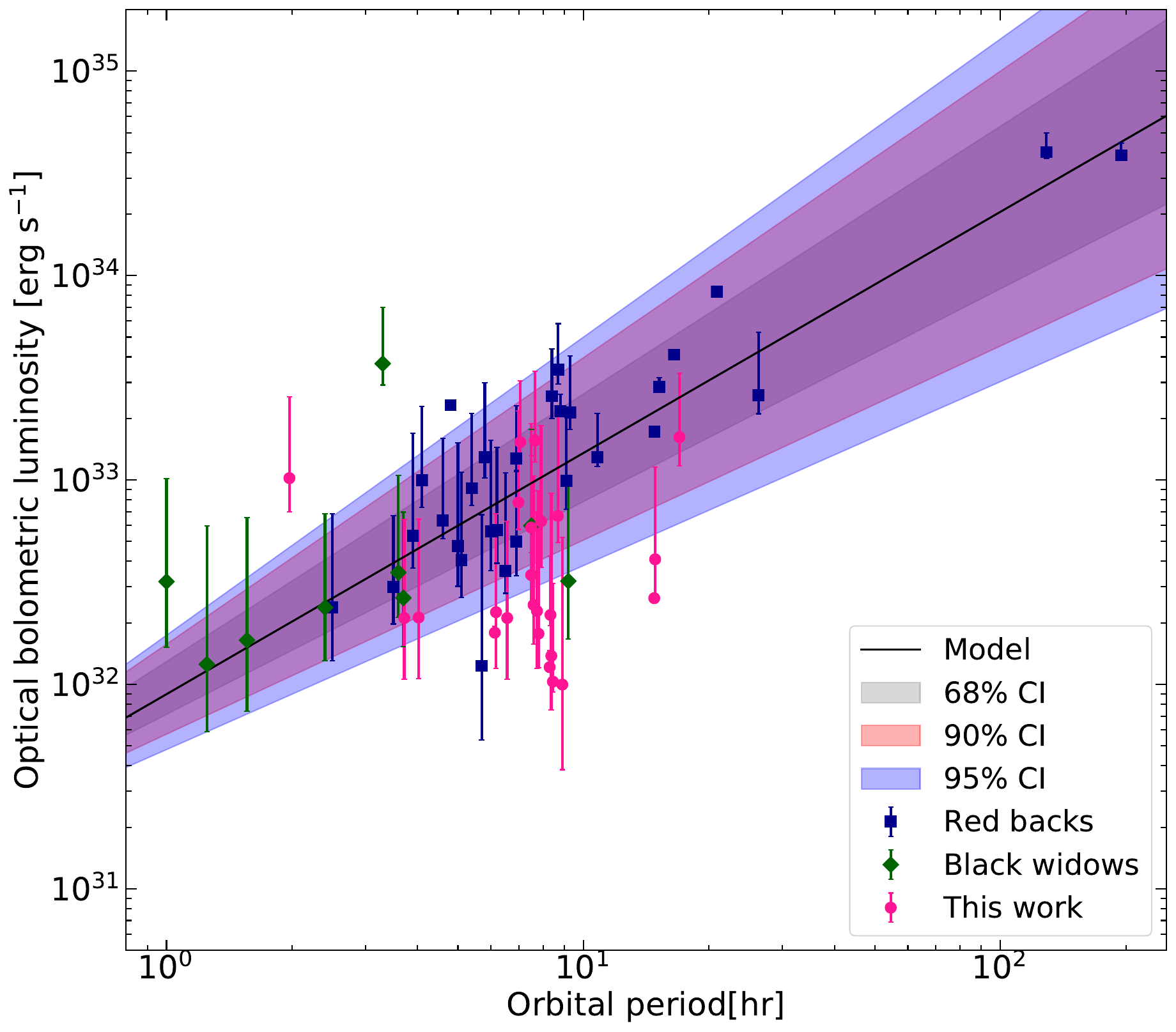}
\caption{The orbital period-optical bolometric luminosity correlation for the known redback sample, the known black widow sample, and the spider Gold sample candidates is depicted. Blue and green dots represent the redback and black widow systems, respectively, from Table 1 in \citet{Koljonen2023}. The pink dots correspond to the spider Gold sample candidates that we have selected based on the Gaia Hertzsprung-Russell diagram. The black solid line represents the best-fit curve derived from our Bayesian MCMC method, and the shaded grey regions indicate the 68 percent, 90 percent, and 95 percent confidence intervals.}
\label{fig:periodluminosity}
\end{center}
\end{figure*}

\section{Discussion} \label{sec:discussion}

\subsection{Radio Observations}

Clark et al. (2023) conducted radio observations of 79 potential gamma-ray pulsar candidates identified from the 4FGL catalog using MeerKAT's L-band receiver (856–1712 MHz). Among these, 22 Spider pulsars (see Figure \ref{fig:goldsample}) from the golden sample were matched with the pulsars detected in the TRAPUM L-band survey. Specifically, 4FGL J1816.7+1749 (with the optical counterpart ZTFJ181628.06+174712.41) was observed with the radio telescope. MeerKAT observations centered on 4FGL were performed within a 5 arcminute scan range. Clark et al. carried out two epochs of observations for 4FGL J1816.7+1749: Epoch 1 (59123.7009) and Epoch 2 (59196.5063). No radio pulse signals were detected near the gamma-ray source in either of the two epochs.

\subsection{X-ray/UV Observations}

We searched for counterparts at other wavelengths for 24 Spider pulsar candidates using the X-ray and UV band catalogs from the HEASARC Browse database \footnote{\url{https://heasarc.gsfc.nasa.gov/db-perl/W3Browse/w3browse.pl}}.
The golden sample was cross-matched with various X-ray telescope archive databases (see Table \ref{tab:Xrayxmatch}), including the Chandra Source Catalog (CSC), XMM-Newton Science Archive's 4XMM-DR13 catalog, the second Swift X-ray Point Source (2SXPS) catalog, Suzaku Wide-Band All-Sky Monitor (WAM) Catalog, IXPE Master Catalog, The All-Sky Catalog of Faint Extreme-Ultraviolet (EUV) Sources, and the eROSITA/eFEDS (Early Data Release site) main catalogue. Additionally, the golden sample was matched with certain UV observational data. The cross-matching results with X-ray/UV showed no counterparts detected in other bands for the golden samples.

\begin{deluxetable*}{cccccc}
\tablenum{3}
\tablecaption{The X-ray catalogs utilized for crossmatching in the present study adhere to the search strategies delineated by \citet{Wang2024} in Table 1. \label{tab:Xrayxmatch}}
\tablewidth{0pt}
\tablehead{
\colhead{Observatory} & \colhead{Catalog} & \colhead{Energy Range} &  \colhead{Median $1\sigma$ Localization}  & \colhead{Sky Coverage} & \colhead{Number of Sources} \\
}
\startdata
ROSAT & 2RXS \citep{Boller:16:2RXS} & 0.1--2.4\,keV & 15.7\arcsec & 100\% & $\texttimes$ \\
XMM & XMMSL2 \citep{Warwick:12:XMMSlewSurvey} & 0.2--12\,keV & 4.2\arcsec & 84\%  & $\texttimes$ \\
Swift-XRT & 2SXPS \citep{Evans:20:2SXPS} & 0.3--10\,keV & 2.7\arcsec & 9.1\%  & $\texttimes$  \\ 
XMM & 4XMM-DR10 \citep{Webb:20:4XMMDR10} & 0.2--12\,keV & 1.5\arcsec & 3.0\% & $\texttimes$ \\ 
Chandra & CSC 2.0 \citep{Evans:10:CSC} & 0.1--10\,keV & 0.6\arcsec & 1.3\% & $\texttimes$ \\
\enddata
\end{deluxetable*}

\begin{deluxetable*}{ccccccccc}
\tablenum{4}
\tablecaption{Estimation of the median, maximum, and minimum values of the distance and the optical bolometric luminosities in the Gaia G band for a golden sample of 24 spider pulsar candidates, as well as the selection of the Color-RPM diagram and the 4FGL 95\% confidence error ellipse. \label{tab:Golddistance}}
\tablewidth{0pt}
\tablehead{
\colhead{ZTF Name} & \colhead{$\rm d_{DR3}$(med)} & \colhead{$\rm d_{DR3}$(min)} & \colhead{$\rm d_{DR3}$(max)} &  \colhead{$\rm L_{G}$(med)}  & \colhead{$\rm L_{G}$(min)}  & \colhead{$\rm L_{G}$(max)} & \colhead{Fermi Elliptical-error} & \colhead{color-$H_{\rm g}$}  \\
\colhead{} & \colhead{kpc} & \colhead{kpc} & \colhead{(kpc)} & \colhead{($\rm 10^{32}~erg~s^{-1}$)} & \colhead{($\rm 10^{32} ~erg~s^{-1}$)} & \colhead{($\rm 10^{32}~erg~s^{-1}$)} & \colhead{Selection} & \colhead{Selection}
}
\startdata
ZTFJ0018+5116 & 3.68139 & 1.97806 & 7.68767 & 15.62 & 3.343 & 18.49 & $\checkmark$ & $\checkmark$  \\
ZTFJ0327+2351 & 2.99584 & 1.46172 & 7.02252 & 7.766 & 2.036 & 14.03  & $\checkmark$ &  \\
ZTFJ0453+2846 & 2.46647 & 1.23952 & 6.13684 & 5.850 & 1.448 & 12.95   & $\checkmark$ & \\
ZTFJ0516+2740 & 1.32575 & 0.88524 & 2.61436 & 2.187 & 0.241 & 2.066  & $\checkmark$ & \\
ZTFJ0524-2617 & 0.67390 & 0.62889 & 0.72587 & 0.015 & 0.011 & 2.641 & $\checkmark$ & \\
ZTFJ0544+2238 & 3.41550 & 1.00671 & 8.27839 & 2.112 & 1.051 & 4.282  & $\checkmark$ & \\
ZTFJ0705+0000 & 3.40345 & 1.23151 & 8.13774 & 6.279 & 2.557 & 12.15  & $\checkmark$ & \\
ZTFJ1701-2231 & 1.95585 & 0.84887 & 5.87068 & 1.770 & 0.567 & 7.092  & $\checkmark$ &\\
ZTFJ1741-1620 & 2.62528 & 1.29631 & 6.43752 & 6.669 & 1.709 & 14.06 & &   \\
ZTFJ1813+2820 & 3.78214 & 2.16069 & 7.55842 & 15.32 & 2.815 & 15.27  & $\checkmark$ & $\checkmark$ \\
ZTFJ1816+1747 & 0.85834 & 0.65632 & 1.24907 & 1.214 & 0.067 & 0.252   &  & $\checkmark$ \\
ZTFJ1819-1103 & 3.47153 & 1.01848 & 8.34721 & 2.112 & 1.055 & 4.166  & $\checkmark$ & \\
ZTFJ1820-1524 & 3.36236 & 1.04258 & 8.19187 & 2.285 & 1.087 & 4.713  & &  \\
ZTFJ1853-0618 & 3.49003 & 1.32498 & 8.20351 & 4.093 & 1.575 & 7.466  & $\checkmark$ & $\checkmark$ \\
ZTFJ1853-0842 & 2.15986 & 1.05234 & 5.79613 & 3.428 & 0.901 & 9.715  & $\checkmark$ &  \\
ZTFJ1859-0710 & 4.65656 & 2.21627 & 9.45013 & 16.19 & 4.446 & 17.16   & $\checkmark$ & \\
ZTFJ1901-0716 & 1.94947 & 0.63152 & 6.41236 & 1.379 & 0.630 & 7.227  & $\checkmark$ & $\checkmark$ \\
ZTFJ1934+1040 & 2.33955 & 0.50250 & 7.15978 & 0.998 & 0.6151 & 4.235  & $\checkmark$ & $\checkmark$ \\
ZTFJ1947+1729 & 0.73643 & 0.60497 & 0.94254 & 1.791 & 0.057 & 0.140   & $\checkmark$ & \\
ZTFJ1957+1233 & 2.34723 & 0.94365 & 6.58299 & 2.448 & 0.875 & 7.972   & & $\checkmark$ \\
ZTFJ2032-0717 & 0.43407 & 0.28864 & 1.05271 & 1.031 & 0.116 & 2.095 & $\checkmark$ & $\checkmark$  \\
ZTFJ2144+7714 & 3.38216 & 1.06338 & 8.20931 & 2.256 & 1.061 & 4.596  & &  \\
\hline
ZTFJ0336+7503 & 3.82517 & 1.34248 & 8.70375 & 2.649 & 1.116 & 4.309  & $\checkmark$ & \\
ZTFJ1838+3224 & 3.41550 & 1.00671 & 8.27839 & 2.124 & 1.056 & 4.305  & $\checkmark$ & $\checkmark$ \\
\enddata
\tablecomments{In the estimation of distances, \emph{Gaia} parallax parameters and \emph{Gaia} \textbf{parallax\_error} were utilized, employing the computational method provided by \citet{Koljonen2023}. A detailed Bayesian-based calculation process is presented in Appendix \ref{sec:appendixa}. For the calculation of optical bolometric luminosities, the computational process outlined in Section \ref{sec:periodluminosity} was applied.  }
\end{deluxetable*}

\subsection{Simbad Astronomical Database}

To confirm whether the Spider pulsar candidates (see Figure \ref{fig:goldsample}) have been previously observed and identified by other research projects, we checked both the golden and variable source samples using the Simbad Astronomical Database. Queries were made using the ICRS, J2000, 2000 coordinate system, with a cross-matching search radius set to within 3 arcseconds. For the golden sample: three sources have been previously observed or reported in earlier studies. ZTFJ172030.59-244114.29, identified as OGLE BLG-DSCT-11935, is a delta Sct Variable. ZTFJ185918.53-071003.00, identified as ATO J284.8271-07.1675, is a Variable Star. ZTFJ214450.81+771437.25, identified as GALEX J214450.6+771436, is a White Dwarf Candidate. Among all variable sources: out of a total of 735 variable source samples, 234 have been reported in previous studies, mainly including: 67 Variable Stars, 3 Pulsating Variables, 28 RR Lyrae Variables, 77 Eclipsing Binaries, 7 Mira Variables, 3 delta Sct Variables, 20 Long-Period Variables, 1 Blazar, 6 Galaxies, 4 BL Lac objects, 5 Quasars, 2 Clusters of Galaxies, 1 Emission-line Star, and 1 X-ray Source. ZTFJ214352.72+660810.25 is marked as an X-ray Source, identified as [SS2009].

\subsection{Comparison to the Birth of the ELMs}

The investigation titled "Birth of the ELMs" represents a comprehensive survey spearheaded by \citet{El-Badry2021}, leveraging data from the Zwicky Transient Facility (ZTF) to identify cataclysmic variables in the midst of mass transfer and those that have recently disengaged from such interactions. These binary systems are hypothesized to be the evolutionary precursors to an array of exotic compact objects, including Extremely Low Mass White Dwarfs (ELM WDs), AM CVn-type systems, and ultra-compact detached binaries. Within the ZTF survey, a subset of evolved CVs and proto-ELM WDs has been discerned, characterized by the presence of a substantial white dwarf in conjunction with a companion star possessing a mass circa 0.15 solar masses. The photometric light curves of these celestial systems bear the hallmarks of ellipsoidal variability, indicative of their close binary nature and with orbital periods constrained to be less than 6 hours. Of particular interest is the observation that the secondary stars in these evolved CVs and recently detached proto-ELM WDs boast effective temperatures frequently surpassing 7000 K, signifying a critical phase in their stellar evolution.

\citet{El-Badry2021}, utilizing data from the ZTF, identified evolved CVs and proto-ELM white dwarfs that exhibit ellipsoidal variability. These sources are located between the main sequence and the white dwarf cooling sequence in the Gaia Hertzsprung-Russell diagram, as delineated by the inequality: $4(G_{\rm BP}-G_{\rm RP}) + 2.7  \leq  M_{\rm G}  \leq  3.25(G_{\rm BP}-G_{\rm RP}) + 9.625$. This region overlaps with the selection range for the spider Gold Sample. To further distinguish between the spider Gold Sample candidates and evolved CVs/proto-ELM white dwarfs, the parameter distributions of the candidates were examined. Figure \ref{fig:cornerplot} shows a corner plot illustrating the correlations between the parameters of known spider pulsars, Gold Sample candidates, and evolved CVs/proto-ELM white dwarfs. It is evident from the plot that the parameter correlations for evolved CVs/proto-ELM white dwarfs are more tightly clustered, whereas the known spider pulsars and Gold Sample candidates exhibit a more dispersed distribution. The parameter distribution of the known spider pulsars and Gold Sample candidates is similar and markedly different from that of evolved CVs/proto-ELM white dwarfs, suggesting that the Gold Sample is more likely to consist of spider pulsar candidates rather than evolved CVs/proto-ELM white dwarfs. Figure \ref{fig:cdf} shows the cumulative distribution functions of the orbital period parameter for known redback, black widow systems, Spider Gold sample candidates, and evolved CVs/proto-ELM white dwarfs. The plot reveals that the Spider Gold sample candidates align with the distribution of known redback and black widow systems and are distinctly different from evolved CVs/proto-ELM white dwarfs.

\subsection{Orbital period-optical luminosity correlation in spider pulsars}   \label{sec:periodluminosity}
 
Without the availability of spectral observational data and additional validation in the radio and X-ray bands, it is exceedingly difficult to ascertain whether these Gold Sample candidates are indeed millisecond pulsars, colloquially referred to as spider pulsars. In this context, we have availed ourselves of the Gaia and ZTF archival data to further explore and extract valuable insights. \citet{Koljonen2023} have put forth a statistical analysis approach to investigate the relationship between the optical and X-ray luminosities characteristic of spider pulsars. Through the estimation of the optical and X-ray luminosities predicated on the parallax distances of the optical counterparts to these spiders, a significant correlation was identified between the optical luminosity and the orbital period of the systems.

In our investigation, we explored the utility of various filters from the \textit{Gaia} mission to assess their efficacy in revealing correlations within our dataset. The results obtained were consistent across different filters, underscoring the robustness of our findings.

Given the established correlations, we proceeded to estimate the dereddened magnitude, leveraging the known distance and orbital period of the celestial objects, without incorporating a bolometric correction. Based on the results of \citet{Koljonen2023}, this estimation is formulated as follows:
\begin{equation}
m_{\mathrm{G}} \approx M_{\odot, \mathrm{bol}} - 2.5 \, \mathrm{log} \frac{L/L_{\odot}}{(d/10)^2} \, ,
\end{equation}
where $L$ is articulated in erg/s, harnessing the least squares regression outcomes delineated in Table \ref{tab:Golddistance}, and $d$ is rendered in parsecs. By adopting $M_{\odot, \mathrm{bol}}$ as 4.74 mag and $L_{\odot}$=3.85$\times10^{33}$ erg/s. Based on the results derived by \citet{Koljonen2023}, here we provide a streamlined formulation of the equation:
\begin{equation}
m_{\mathrm{G}} \approx 16.26 + 2.5 \, \mathrm{log} \frac{d_{\mathrm{kpc}}^2}{L_{33}} \, ,
\end{equation}
with \( d_{\mathrm{kpc}} \) expressed in kiloparsecs and $L_{33}$ in units of $10^{33}$ erg/s.

By integrating the luminosity surmised from the least squares regression, predicated on the \textit{Gaia} DR3 distances of the parallax-selected sample, we are positioned to approximate the anticipated \textit{Gaia} G-band magnitude. This approximation is contingent upon the distance and orbital period, and is mathematically represented as \citet{Koljonen2023}:
\begin{equation}
m_{\mathrm{G}} \approx 19.3(\pm 0.7) + 2.5 \, \mathrm{log} \frac{d_{\mathrm{kpc}}^2}{P_{\mathrm{hr}}^{1.34(\pm0.21)}} \, .
\end{equation}

This formulation serves as a valuable tool for astronomers, enabling the prediction of the G-band magnitude for similar astronomical entities, thereby fostering a deeper comprehension of their intrinsic luminosities and distances (see Table \ref{tab:Golddistance}). The precision of this approximation is subject to the uncertainties associated with the regression analysis, as reflected in the error terms presented.

Figure \ref{fig:periodluminosity} illustrates the correlation between the orbital period and optical luminosity for known redback and black widow systems, as well as the Spider Gold sample candidates. The redback and black widow systems are derived from Table 1 of \citet{Koljonen2023}. We have recalculated the Gaia G-band luminosity for these sources, employing geometric distance estimates based on Gaia DR3 parallaxes. The distance estimation method using Gaia parallax parameters adheres to the approach detailed by \citet{bailerjones21}. \citet{Koljonen2023} have previously determined the best-fitting linear least squares regression for the sample, which is $\rm log(L/erg s^{-1}) = 1.34(\pm 0.21) \times log(P/h) + 31.78(\pm 0.28)$. Utilizing the most recent distance parameters, we have refitted the orbital period to optical luminosity data for the known spider millisecond pulsars via the Markov Chain Monte Carlo method, yielding the best-fit model as $\rm log(L/erg s^{-1}) = 1.17(_{-0.133}^{
+0.135}) \times log(P/h) + 31.98(_{-0.132}^{+0.126})$. The figure \ref{fig:periodluminosity} delineates the 68\%, 90\%, and 95\% confidence intervals for the model. Employing an identical approach, we have estimated the luminosity for the Spider Gold sample candidates and depicted the data points alongside their uncertainties. Notably, three of the Gold sample candidates are situated within the black widow distribution region, with the remainder falling within the redback distribution area. These findings indicate a strong likelihood that the Spider Gold sample candidates are associated with redback systems, contingent upon spectral confirmation.

\section{Conclusion} \label{sec:conclusion}

In this paper, we have systematically investigated redback and black widow candidates by combining the 4FGL-DR3 unassociated sources with ZTF time-domain variability data. The search for and authentication of spider pulsars contribute to advancing our understanding of these enigmatic systems. The new spider pulsar candidate list is crucial for testing and refining the "Recycling" theory of millisecond pulsar formation. These systems provide a unique observational sample for studying the complex interplay of accretion, irradiation, and stellar evolution in compact binaries. Here, we distill the essence of our findings and their implications for astrophysics.

Our discovery demonstrates the advantage of ZTF data in detecting spider pulsar binary systems. We have developed a large-scale time-domain data processing pipeline using wget and the Lomb-Scargle Periodogram algorithm, capable of searching for optical counterparts of Fermi gamma-ray sources within a 6-arcminute range and generating a wealth of light curve data. For the test sample, a total of 42 known RB and BW systems were identified, with 12 light curves exhibiting periodic variations, of which 5 belong to RB systems and 7 to BW (or BW candidate) systems. The identification of periodic variations in the light curves of known spider pulsars in the test sample confirms the reliability of our efficient search strategy.

Subsequently, we systematically explored the parameter space of 2,179 sources from the 4FGL-DR3 unassociated sources catalog and cross-matched them with optical counterparts from ZTF, finding 1,351 gamma-ray sources within the ZTF survey area. Through the identification and classification of variable light curve types, we discovered 1,566 optical counterparts with periodic variations, yielding a rich set of 194 EV-type binary star candidates. Additionally, we identified two Irradiation Binary samples: ZTF J0336+7502 and ZTF J1838+3224, both of which have been confirmed as black widow candidates in other studies. 
Utilizing the distribution patterns of known spider pulsars in the Gaia Hertzsprung-Russell diagram, we further refined our selection, ultimately determining a Gold Sample of 24 spider pulsar candidates. Subsequent application of the 4FGL 95\% confidence error ellipse revealed that 19 candidates met the selection criteria. Further refinement using RPM diagrams allowed us to conduct a stringent screen, ultimately confirming that 9 candidates passed the rigorous selection standards.

The candidates identified in this study require follow-up observations across the electromagnetic spectrum, particularly in radio and X-ray bands, to confirm their nature and further characterize their properties. We matched the Gold Sample of spider pulsars with Radio, X-ray/UV data and found no observational confirmation in these bands. Additionally, we compared the candidates with the Simbad Astronomical Database.

Our refined Gold Sample is poised to enhance the population of known spider pulsars, offering a more comprehensive view of the distribution and characteristics of black widows and redbacks across the galaxy. The correlation between optical luminosity and orbital period derived from our analysis provides a robust predictive tool for identifying similar systems, guiding future observational campaigns. Furthermore, we found that the Gold Sample exhibits distinct differences from the progenitors of extremely low-mass white dwarfs (ELM WDs), such as evolved CVs and proto-ELM WD systems. Future studies will require a combination of photometry, spectroscopy, and astrometry to further confirm their physical properties and unravel the evolutionary paths of these compact binaries.

The development of more sophisticated data processing techniques and period-finding algorithms is essential for managing the vast amounts of data from current and upcoming sky surveys, ensuring the accurate identification of variable sources. In the classification of light curves, we will further utilize machine learning algorithms to assist in classification, improving search efficiency.

\acknowledgments

We acknowledge the science research grants from the China Manned Space Project. CYL, WJH, HWY and PHT are supported by the National Natural Science Foundation of China (NSFC) under grant 12273122. LLR and JML gratefully acknowledge support from the NSFC through grant No. 12233013.

This work has made use of data from the European Space Agency (ESA) mission Gaia (\url{https://www.cosmos.esa.int/gaia}), processed by the Gaia Data Processing and Analysis Consortium (DPAC, \url{https://www.cosmos.esa.int/web/gaia/dpac/consortium}). Funding for the DPAC has been provided by national institutions, in particular the institutions participating in the Gaia Multilateral Agreement. 
The ZTF data used in this study can be accessed from the ZTF Science Data System (ZSDS) \citep{Masci2019} housed at NASA/IPAC Infrared Science Archive (IRSA ZTF Lightcurve Queries API, \url{https://irsa.ipac.caltech.edu/docs/program_interface/ztf_lightcurve_api.html}).

\vspace{5mm}
\facilities{PO:1.2m (ZTF), Gaia, ADS, CDS.}
\software{ Astropy \citep{Astropy2013}, matplotlib \citep{Hunter2007}, NumPy \citep{Harris2020}, Pandas \citep{McKinney2010}, SciPy \citep{Virtanen2020} and the IPython package \citep{Perez2007}}.


\bibliography{sample63}{}
\bibliographystyle{aasjournal}

\appendix

\section{Gaia Distance Estimation} \label{sec:appendixa}
In order to ascertain the optical luminosity of the pulsar sample for the Spider Pulsar, it is imperative to determine the Gaia distances of these celestial bodies. In a previous work, \citet{Koljonen2023} delineated a procedure for the computation of Gaia distances for the Spider Pulsars within Appendix A, titled "GAIA GEOMETRIC DISTANCE." We have elected to employ this methodology and offer an exhaustive exposition of it.
We calculate the posterior probability density function for the distance to each source, which is derived from the product of the likelihood and the prior, as described by \citet{bailerjones21} and \citet{Koljonen2023}. The posterior is given by:
\begin{equation}
    P (d | \omega, \sigma_{\omega}, p) = P (\omega |d, \sigma_{\omega}) P (d | p), 
\end{equation}
where $d$ represents the distance, $\omega$ is the parallax, 
$\sigma_{\omega}$ is the parallax error, and $p$ denotes a HEALpixel number. We have considered two distinct priors for distance estimation: one based on stellar populations as employed by \citet{bailerjones21}, and another derived from the pulsar distribution modeled by \citet{lorimer06}. It is important to note that while the pulsar-based prior may be more suitable for spider systems, its accuracy is highly dependent on the assumed distribution of free electrons within the Galaxy. \citet{lorimer06} utilized the distribution from \citealt{cordes02}, which, combined with the significantly lower statistics of pulsars compared to stars, can impact the accuracy of the distribution estimate.

The prior used by \citet{bailerjones21} is a three-parameter generalized gamma distribution that varies with the HEALpixel. Each sky direction has unique prior parameters ($\alpha$, $\beta$, L), which are detailed in the online supplementary material of \citet{bailerjones21} and \citet{Koljonen2023}. The prior is expressed as:
\begin{equation}
     P (d | p) = 
     \begin{cases}
       \frac{1}{\Gamma(\frac{\beta+1}{\alpha})} \frac{\alpha}{L^{\beta+1}} d^{b}  e^{-(d/L)^{\alpha}} & \mathrm{if} \, d \geq 0 \\
       0 & \text{otherwise.}
     \end{cases}
\end{equation}

The Galactic distribution of pulsars, as characterized by \citet{lorimer06} and transformed into an Earth-based coordinate system by \citet{verbiest12}, is represented by a double-exponential function incorporating both radial and distance-above-the-Galactic-plane components \citep{Koljonen2023}. The prior in this Earth-based system is given by:
\begin{equation}
    P (d | G_{\mathrm{b}}, G_{\mathrm{l}}) \propto R^{1.9} \mathrm{exp} \Bigg[ -\frac{|z|}{E} - 5\frac{R-R_{0}}{R_{0}}\Bigg]d^2 ,
\end{equation}

\noindent where 

\begin{equation}
    z(d, G_{\mathrm{b}}) = d\sin{G_{\mathrm{b}}} \, ,
\end{equation}

\noindent and

\begin{equation}
    R(d, G_{\mathrm{b}}, G_{\mathrm{l}}) = \sqrt{R_{0}^{2} + (d\cos{G_{\mathrm{b}}}) - 2 R_{0} d \cos{G_{\mathrm{b}}} \cos{G_{\mathrm{l}}} } \, .
\end{equation}

The constants $R_{0}$ and $E$ represent the distance to the Galactic center and the scale height, respectively, with values of 
$R_{0} = 8.12$ kpc and $E = 0.4$ kpc \citep{Koljonen2023}. For Gaussian parallax uncertainties, the likelihood is defined as:
\begin{equation}
    P (\omega |d, \sigma_{\omega}) = \frac{1}{\sqrt{2\pi}\sigma_{\omega}} \mathrm{exp} \Bigg[-\frac{1}{2\sigma_{\omega}^{2}} \Bigg(\omega-\omega_{\mathrm{zp}}-\frac{1}{d} \Bigg)^{2} \Bigg],  
\end{equation}
where  $\omega_{\mathrm{zp}}$ is the parallax bias.

\section{\emph{4FGL} error ellipse and ZTF search area sky map}

\begin{figure*}[htpb]
\begin{center}
\includegraphics[width=0.9\textwidth]{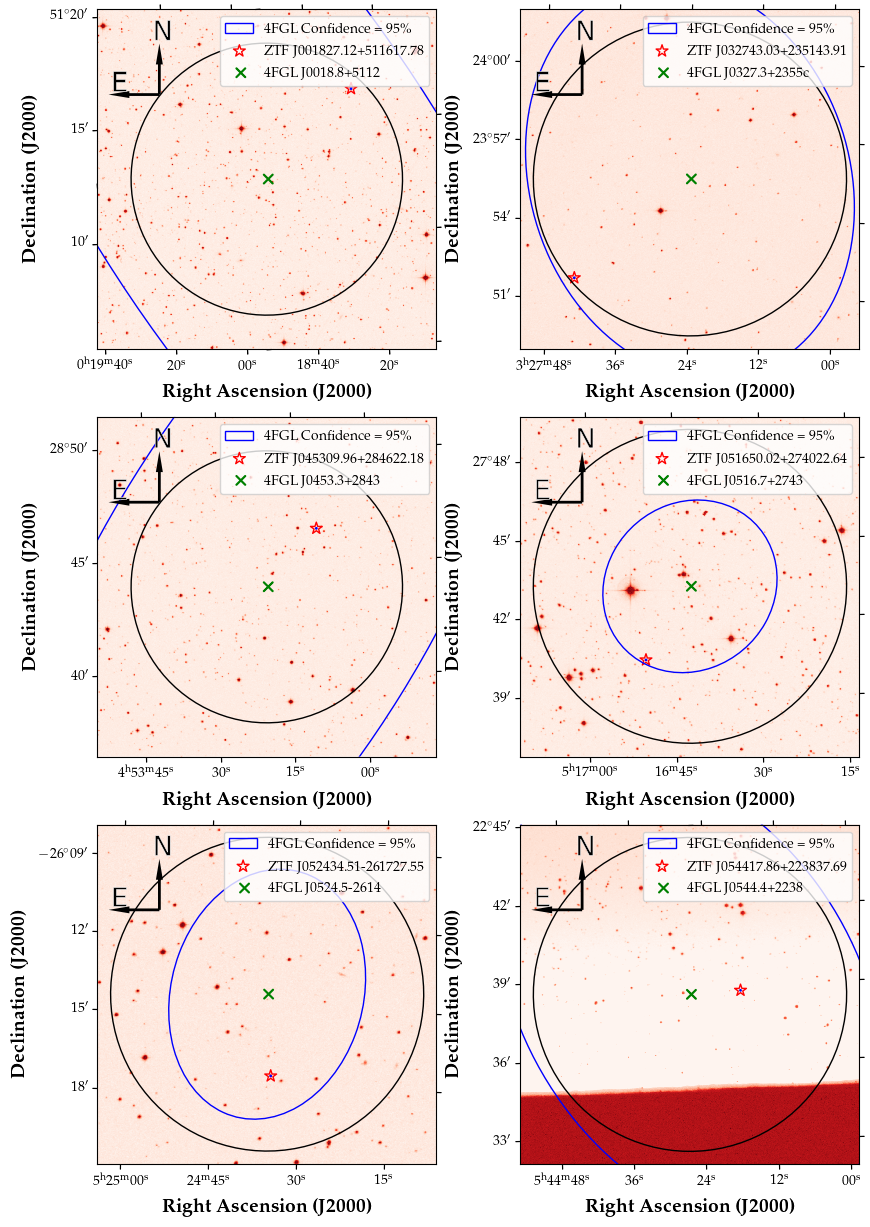}
\caption{The sky maps of the 24 the spider Gold sample in the search sample. The blue ellipses represent the 4FGL 95\% confidence error ellipses, the black circles indicate the search area within 6 arcminutes in the ZTF variability data. The green crosses mark the central coordinates of the Fermi gamma-ray sources, and the red pentagons denote the optical counterparts in the ZTF. The background data for the sky maps are sourced from the POSS-1 red band image, accessed through the Virtual Observatory archives.}
\label{fig:possimage3}
\end{center}
\end{figure*}

\begin{figure*}[htpb]
\begin{center}
\includegraphics[width=0.9\textwidth]{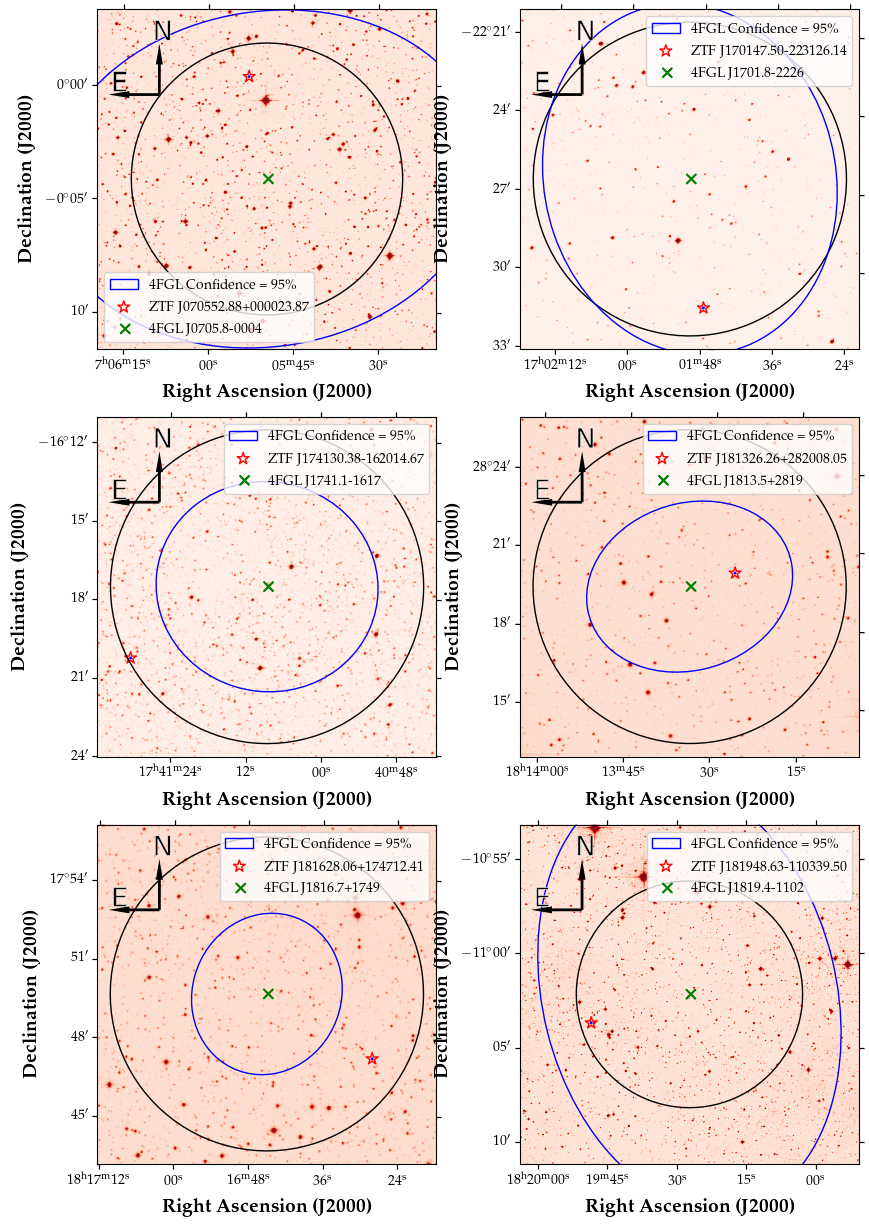}
\caption{POSS-1(red) Image of the spider Gold sample. Same as in Figure \ref{fig:possimage3} (continued).}
\label{fig:possimage4}
\end{center}
\end{figure*}

\begin{figure*}[htpb]
\begin{center}
\includegraphics[width=0.9\textwidth]{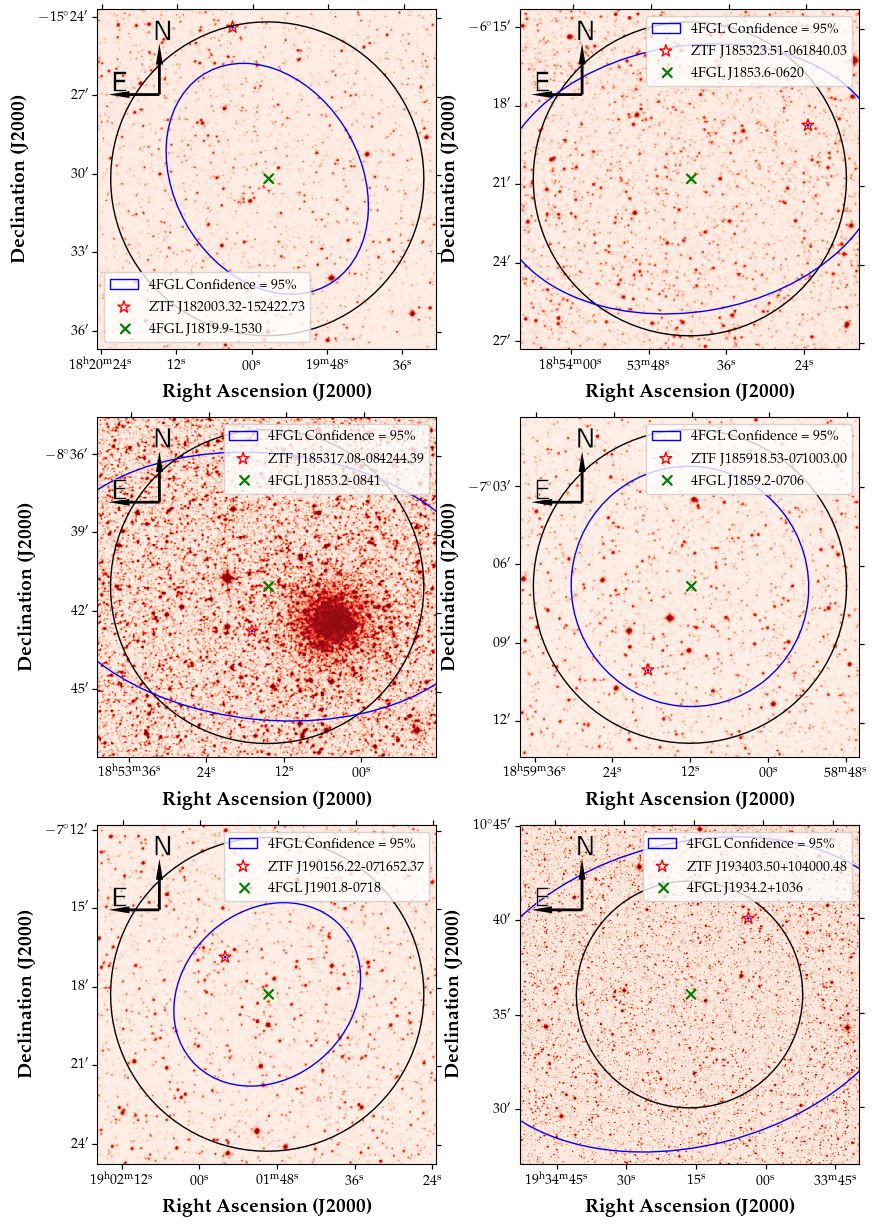}
\caption{POSS-1(red) Image of the spider Gold sample. Same as in Figure \ref{fig:possimage3} (continued).}
\label{fig:possimage5}
\end{center}
\end{figure*}

\begin{figure*}[htpb]
\begin{center}
\includegraphics[width=0.9\textwidth]{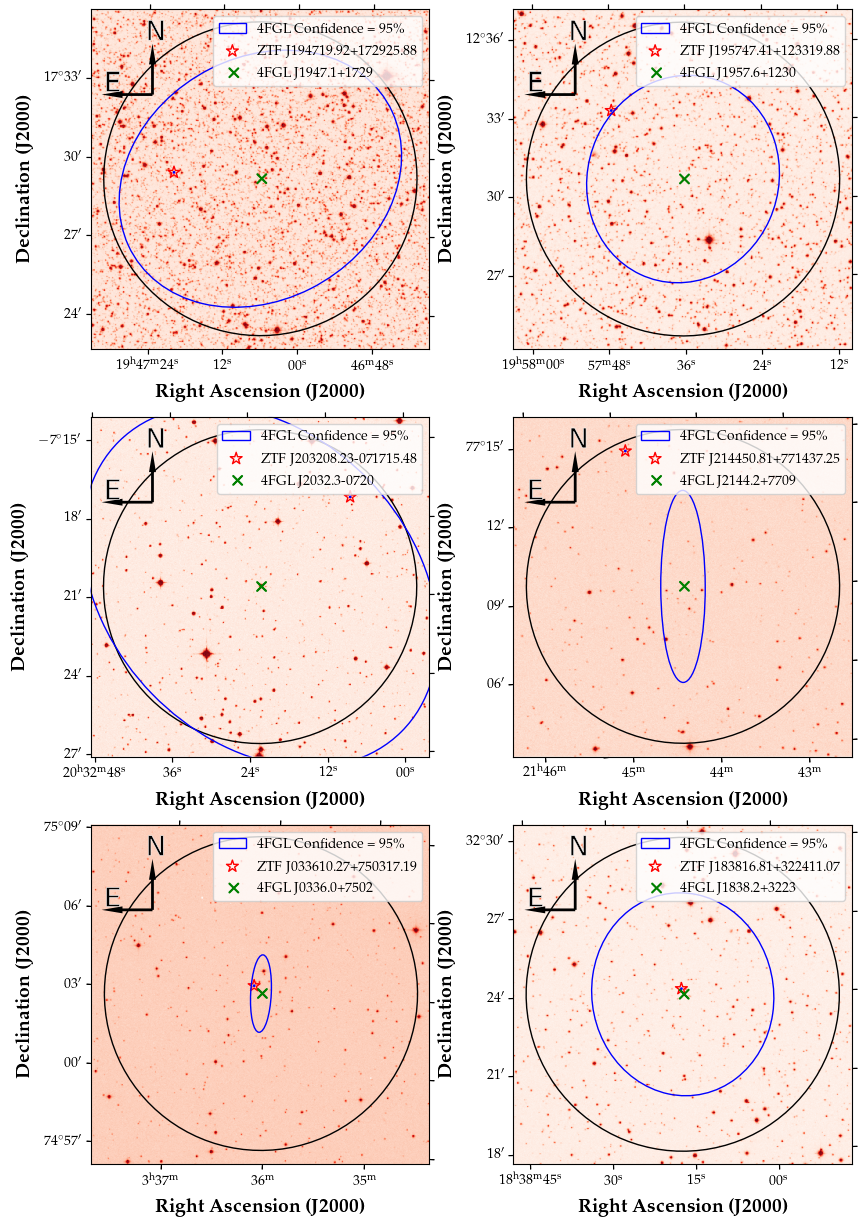}
\caption{POSS-1(red) Image of the spider Gold sample. Same as in Figure \ref{fig:possimage3} (continued).}
\label{fig:possimage6}
\end{center}
\end{figure*}

Below, we illustrate the 4FGL error ellipses and the ZTF search area sky maps for 24 candidate golden samples of spider pulsars. Figures \ref{fig:possimage3} through \ref{fig:possimage6} depict the sky maps for the golden sample, with the central point of each map indicating the location of an unidentified Fermi gamma-ray source. The red pentagon symbol represents the ZTF optical counterpart, the blue ellipse corresponds to the 95\% confidence 4FGL error ellipse, and the black circle delineates the 6-arcminute search area of the ZTF data. The background data for the sky maps are sourced from the POSS-1 red band image, accessed through the Virtual Observatory (VO) archives.

\end{CJK*}
\end{document}